\documentclass[aps,prd,two column]{revtex4-2}

\usepackage{amsmath}
\usepackage{ulem}
\usepackage{color}
\usepackage{graphicx}
\usepackage{epsfig}
\usepackage{bm}
\usepackage{hyperref}
\usepackage{natbib}
\usepackage{caption}
\usepackage{subcaption}

\def\be {\begin{equation}}
\def\ee {\end{equation}}
\def\nn {\nonumber}
\def\bea {\begin{eqnarray}}
\def\eea {\end{eqnarray}}

\begin{document}
\title{Magneto-Thomson and transverse Thomson effects in an interacting hadron gas in the presence of an external magnetic field}
 
\author{Kamaljeet Singh}
\author{Kshitish Kumar Pradhan}
\author{Raghunath Sahoo}
\email{Corresponding Author: Raghunath.Sahoo@cern.ch}
\affiliation{Department of Physics, Indian Institute of Technology Indore, Simrol, Indore 453552, India}

\date{\today}

\begin{abstract}
The universality of electric charge as a quantum number allows thermoelectric properties to manifest across diverse systems, starting from a hot quantum chromodynamic matter in heavy-ion collisions at a high energy scale to semiconductors in condensed matter systems at a low energy scale.
In this work, we explore the emergence of magneto-transport phenomena, specifically the magneto-Thomson and transverse Thomson effects, in a hot and dense hadronic medium produced in relativistic heavy-ion collisions at the Relativistic Heavy Ion Collider and Large Hadron Collider energies. These phenomena arise due to the combined influence of temperature gradients and non-zero baryon chemical potential, particularly in the presence of an external magnetic field. Using the relativistic Boltzmann transport equation within the relaxation time approximation, we analyze the behavior of the hadronic medium considering different frameworks of hadron resonance gas models. The presence of external magnetic fields breaks the isotropy of the thermoelectric transport coefficient matrix, giving rise to new components of the Thomson coefficient, namely, magneto-Thomson and transverse Thomson coefficients. For the first time, we estimate the magneto-Thomson and transverse Thomson coefficients, which originate from the temperature dependence of the magneto-Seebeck coefficient and Nernst coefficient, respectively, in hadron gas under the influence of a static and time-varying magnetic field. Our findings provide a novel perspective on the higher-order thermoelectric properties of the hot and dense hadronic medium in the context of heavy-ion collisions.
\end{abstract}

\maketitle

\section{Introduction}
Scientific curiosity about the early universe has led to groundbreaking progress in experimental physics, particularly in the study of strongly interacting matter governed by quantum chromodynamics (QCD). Facilities like the Large Hadron Collider (LHC) and the Relativistic Heavy Ion Collider (RHIC) serve as modern laboratories for probing the thermodynamic and transport properties of strongly interacting matter, such as the quark-gluon plasma (QGP) and hot hadron gas~\cite{Busza:2018rrf, Brewer:2019oha}. This QGP phase of matter forms at very high temperatures and/or baryon chemical potentials and exists for a short lifetime with an order of a few fermi. As the QGP expands and cools rapidly, it undergoes hadronization, eventually leading to a dense hadron gas that freezes out into detectable subatomic particles. Understanding the properties of this hadronic phase is essential for reconstructing the full evolution of the collision and probing the characteristics of dense nuclear matter. During this evolution, temperature gradients naturally develop in space and time due to the rapid expansion and cooling of the medium from the central collision zone to outward. These spatial and temporal temperature variations significantly influence the dynamics of the system. In the presence of conserved charges, such as baryon number or electric charge, these temperature gradients can drive charge carriers to diffuse along these gradients, leading to the development of thermoelectric currents~\cite{book}. These currents lead to the study of thermal and electrical conductivity, particularly relevant in the context of heavy-ion collisions~\cite{Singh:2023pwf, Singh:2023ues}. In the hadron resonance gas (HRG) medium, the thermal, and charge transport and the interplay between these two gives rise to thermoelectric coefficients, such as the Seebeck~\cite{PRB98,cai,PhysRevB.105.235116} and Thomson~\cite{Young1924,MacDonald2006,Morrison2020} coefficients, which characterize the response of the medium to thermal forces. 
 The leading-order coefficient, such as the Seebeck coefficient, in particular, characterizes the ability of a conducting medium to convert thermal gradients into electric potential and is defined under open-circuit conditions where the net electric current is zero~\cite{Singh:2024emy}. For a number of materials, the Seebeck coefficient is not spatially uniform but instead varies with temperature, giving rise to higher-order thermoelectric effects. When an electric current passes through such a gradient, the system exhibits the Thomson effect, quantified by the Thomson coefficient. This coefficient represents the amount of net heat absorbed or emitted per unit current per unit temperature gradient, with its magnitude and sign dependent on the intrinsic properties of the medium and the directionality of both current flow and thermal gradient~\cite{Singh:2025rwc}. In the domain of high-energy physics, the leading-order thermoelectric properties of a hot and dense hadron gas have been explored through several theoretical frameworks, including relativistic kinetic theory~\cite{Bhatt:2018ncr, Das:2020beh, Das:2021qii}, hydrodynamic modeling~\cite{Greif:2017byw, Prakash:1993bt, Kadam:2014xka}, and lattice QCD simulations~\cite{Meyer_2011, Kaczmarek_2022}. Understanding these coefficients provides insights into non-equilibrium complex transport phenomena and the microscopic properties of hot QCD matter under extreme conditions.

The dynamics of thermoelctricity is affected by the presence of a nonzero magnetic field because of the experience of the Lorentz force by charge carriers. Peripheral heavy-ion collisions are capable of generating extremely strong magnetic fields as a result of the relativistic motion of spectator protons. In such events, the magnetic field strength can peak around $m_\pi^2 \sim 10^{18}$~G for Au-Au collisions at RHIC and reach up to $15~m_\pi^2$ in Pb-Pb collisions at the LHC~\cite{Wang2021,MCINNES2016173}. These fields tend to dissipate almost instantly in a vacuum, but the situation changes when a thermalized conducting medium, such as the quark-gluon plasma, is present. Due to its finite electrical conductivity, the medium supports the induction of secondary magnetic fields aligned with the initial external field, as described by Faraday’s law~\cite{DEY2023122654}. This results in a slower decay of the magnetic field, allowing it to play a significant role in influencing the complete evolution of the fireball, particularly its thermodynamic and transport properties. In the hadronic phase, implementing a time-varying magnetic field requires an evolving description of the medium, which is not inherently present in the static thermodynamic framework of the HRG model. Unlike the QGP phase, where hydrodynamic profiles can be incorporated to describe the space–time evolution and hence the decay of the magnetic field, the HRG model in its conventional form lacks such dynamic input. However, in the hadronic phase, the space-time evolution of the matter and hence of the electromagnetic fields can be incorporated by means of kinetic theory. Therefore, we adopt both static and dynamical magnetic field pictures in the HRG phase. We use two representative strengths for the static picture ($eB = 0.1$ and $ ~1.0~ m_\pi^2$)~\cite{Ayala:2018zat} and the dynamical picture with magnetic decay parameter $\tau_B$ = 6 fm, to qualitatively mimic the possible range of field strengths during the late stages of the collision.
The influence of magnetic fields on the conductivity of QCD medium, both electrical and thermal, has been extensively addressed in the literature~\cite{Singh:2023pwf,PhysRevD.106.034008,PhysRevD.104.094037}. Under such magnetic field conditions, leading-order thermoelectric transport becomes directionally dependent, giving rise to the magneto-Seebeck and Nernst coefficients. Unlike the Seebeck coefficient, which responds to longitudinal temperature gradients alone, the Nernst coefficient is a transverse response that emerges only in the presence of magnetic fields. It quantifies the generation of an electric current perpendicular to both the thermal gradient and the magnetic field~\cite{Singh:2024emy}. Previous studies have analyzed this behavior for the hadronic medium using hadron resonance gas models~\cite{Bhatt:2018ncr, Das:2020beh}, while the study~\cite{Zhang:2024htn} explores the impact of magneto-thermoelectric effects on conserved charge diffusion. The thermoelectric behavior of the QGP has also been explored in both isotropic~\cite{Abhishek:2020wjm} and anisotropic~\cite{Kurian:2021zyb,Zhang:2020efz} media, largely within perturbative QCD formalisms, which are valid at very high temperatures. The temperature dependence of the leading-order thermoelectric coefficients in a magnetic field gives rise to the higher-order thermoelectric coefficients. In our previous study~\cite{Singh:2025rwc}, we calculated the Thomson coefficient of interacting hot hadron gas, which originates due to the temperature dependence of the Seebeck coefficient. The Thomson effect can be understood as originating from the simultaneous operation of the Seebeck and Peltier effects~\cite{Kimura2024}, and appears when the medium has a temperature dependence of the Seebeck coefficient. If the charge current flows through the spatial gradient of the Seebeck coefficient, the self-induced Peltier effect modulates the temperature in response to the charge current~\cite{PhysRevLett.125.106601}.
In the current study, we calculate the Thomson effect for hardon gas in the presence of a magnetic field. The presence of a magnetic field creates an anisotropy in the medium and introduces magneto-Thomson and transverse Thomson coefficients. For the first time, we calculate the magneto-Thomson coefficient and transverse Thomson coefficient for the hot and dense hadron gas. The transverse Thomson effect originates due to the presence of the Nernst effect in the presence of a magnetic field~\cite{takahagi}. In the absence of a magnetic field, the magneto-Thomson coefficient reduces to the Thomson coefficient, but the transverse Thomson coefficient vanishes completely. 
In principle, strong magnetic fields, such as those generated in the early stages of non-central heavy-ion collisions at LHC energies, can modify hadronic properties through mechanisms such as Landau quantization, which may in turn influence hadron-hadron interactions. It should be noted that, besides Landau quantization, strong magnetic fields can also modify hadron–hadron interactions by affecting the interaction vertices and propagators, particularly for charged hadrons~\cite{Kharzeev:2013jha, Bali:2017ian, Loewe:2017kiw, Ayala:2015lta}. Such modifications could in principle influence scattering cross-sections and relaxation times, thereby altering transport coefficients. A detailed quantum-field-theoretic treatment of these effects lies beyond the scope of the present kinetic-theory framework. Here, the effect of the magnetic field enters into the transport equations through the cyclotron frequency of individual hadrons, while the hadronic scattering cross-sections are taken to be field-independent within the hard‑sphere approximation. The effects arising from Landau quantization for the charged hadrons are also explicitly considered in this study. 

To study the properties of the hot and dense hadronic medium created in the later stages of heavy-ion collisions, the ideal hadron resonance gas (IHRG) model has been widely used. It successfully describes thermodynamic quantities predicted by lattice QCD (lQCD) up to temperatures of approximately \(T \approx 150~\mathrm{MeV}\), beyond which hadron melting leads to notable deviations from lQCD~\cite{Bellwied:2013cta, Bellwied:2017ttj}. Unlike lQCD, which becomes computationally challenging at large baryon chemical potential due to the fermion sign problem~\cite{Borsanyi:2013bia, HotQCD:2014kol}, the IHRG model remains useful for exploring hadronic matter at high baryon density. However, its inability to capture higher-order fluctuations of conserved charges, well-predicted by lQCD—has driven the development of improved models~\cite{Bazavov:2013dta, Bazavov:2017dus, Borsanyi:2018grb}. Among these, the excluded volume HRG (EVHRG) model introduces a finite size for hadrons, accounting for repulsive interactions via an excluded volume correction. This enhancement leads to better agreement with both thermodynamic observables and conserved charge fluctuations when compared to lQCD data~\cite{Andronic:2012ut, Kadam:2015xsa, Kadam:2017iaz, Pal:2020ucy}. Another model, the repulsive mean-field HRG (RMFHRG) model~\cite{Kapusta:1982qd, Olive:1980dy}, incorporates a mean-field potential to model inter-hadron repulsion. Recently, the van der Waals HRG (VDWHRG) model~\cite{Vovchenko:2017cbu, Vovchenko:2016rkn} has gained attention for including both attractive and repulsive forces, making it one of the most realistic HRG extensions. It shows excellent agreement with lQCD predictions~\cite{Vovchenko:2016rkn, Samanta:2017yhh} and also predicts a first-order liquid-gas phase transition at high baryon densities~\cite{Vovchenko:2015vxa}, offering valuable insights into dense nuclear matter.

In this work, we provide detailed calculations of the magneto-Thomson and transverse Thomson coefficients of the hot and dense hadronic medium in the presence of an external magnetic field. We use four different formalisms, namely, the IHRG, EVHRG, RMFHRG, and VDWHRG, to estimate the higher-order thermoelectric coefficients. This paper is organized in the following manner. Section (\ref{sec: Formalism}) briefly introduces different HRG models, following which we present the derivation of both leading-order and higher-order thermoelectric coefficients in the presence of an external magnetic field. Section (\ref{results}) discusses the results in detail, and finally, we summarize our study in Sec. (\ref{summary}) with potential future directions.
 
\label{intro}
\section{Formalism}\label{sec: Formalism}
The current section highlights the calculations for both magneto-Thomson and transverse Thomson coefficients in the presence of an external magnetic field. Here, we solve the relativistic Boltzmann transport equation (RBTE) under relaxation time approximation (RTA). We discuss the hadron resonance gas model along with its extended versions that we use to calculate the higher-order thermoelectric coefficients of hot and dense hadron gas. Throughout the calculations, we have taken all the hadrons and their resonances up to the mass cut-off of 2.6 GeV from the particle data group~\cite{ParticleDataGroup:2008zun}.

\subsection{Hadron resonance gas model}
The Hadron Resonance Gas model is a theoretical framework used in high-energy nuclear physics to describe the thermodynamic properties of strongly interacting matter in the hadronic phase. The HRG model assumes that the hadronic matter is in thermal equilibrium, meaning that the temperature, chemical potential, and other thermodynamic variables are well-defined throughout the system. The various HRG models are discussed here.

\subsubsection{Ideal hadron resonance gas model}
The ideal HRG model consists of a system with volume $V$ having non-interacting pointlike hadrons and resonances, for which the grand canonical partition function can be written as~\cite{Pradhan:2022gbm},
\begin{equation}
\label{eq1}
ln Z^{id}_i = \pm \frac{Vg_i}{2\pi^2} \int_{0}^{\infty} k_i^2 dk_i\ ln\{1\pm \exp[-(\omega_i-\mu_i)/T]\}.
\end{equation}
Here, $g_i$ is the degeneracy factor. The quantities $k_i$, $m_i$, and $\omega_i = \sqrt{k_i^2 + m_i^2}$ represents the momentum, mass, and energy of the $i{\rm th}$ hadron species, respectively. The $\pm$ sign corresponds to fermions (upper) and bosons (lower). Considering a simplistic case of vanishing charge and strangeness chemical potential, the total chemical potential of the $i$th hadronic species, $\mu_{i}$, is equal to the baryon chemical potential ($\mu_B$) and is given by 
\begin{equation}
\label{eq2}
\mu_i = \beta_i\mu_B ,
\end{equation}
where $\beta_i$ denotes the baryon number of the $i{\rm th}$ hadron. The pressure $P_i$, energy density $\varepsilon_i$, and number density $n_i$ can now be obtained from the partition function, given as,
\begin{align}
\label{eq3}
P^{id}_i(T,\mu_i) &= \pm \frac{Tg_i}{2\pi^2} \int_{0}^{\infty} k_i^2 dk_i\ ln\{1\pm \exp[-(\omega_i-\mu_i)/T]\}, \\
\varepsilon^{id}_i(T,\mu_i) &= \frac{g_i}{2\pi^2} \int_{0}^{\infty} \frac{\omega_i\  k_i^2 dk_i}{\exp[(\omega_i-\mu_i)/T]\pm1},\\
n^{id}_i(T,\mu_i) &= \frac{g_i}{2\pi^2} \int_{0}^{\infty} \frac{k_i^2 dk_i}{\exp[(\omega_i-\mu_i)/T]\pm1}.
\end{align}

\subsubsection{Excluded volume hadron resonance gas model}
The EVHRG model includes the repulsive interaction among the hadrons by considering a finite hardcore size of the hadrons. Therefore, the system volume ($V$) in the model is now replaced by the available volume ($V_{avl}$) that mimics the short-range repulsive interaction among the hadrons. 
\begin{equation}
    V_{avl} = V-\sum_i N_i\mathrm{b}_i,
\end{equation}
where the $N_i$ is number of $i$th hadron and $b_i$ (=16/3 $\pi r_i^3$) is the excluded volume. The $r_i$ is the hardcore radius of each hadron. The pressure in the EVHRG model is then given by~\cite{Rischke:1991ke, Bhattacharyya:2013oya}

\be
P^{EV}(T,\mu)=P^{\text{id}}(T,\mu^{*}),
\label{prexcl}
\ee
where $\mu^{*}=\mu-\mathrm{b}P^{EV}(T,\mu)$ is an effective chemical potential. Other thermodynamic quantities can be readily obtained from Eq.~(\ref{prexcl}) by taking appropriate derivatives. The number density and energy density, respectively, can be written as~\cite{Rischke:1991ke}
\be
n^{EV}(T,\mu)=\sum_{i}\frac{n^{id}_{i}(T,\mu^{*})}{1+\sum_{i}\mathrm{b}_{i}n_{i}^{id}(T,\mu^{*})},
\ee
\be
\varepsilon^{EV}(T,\mu)=\sum_{i}\frac{\varepsilon^{id}_{i}(T,\mu^{*})}{1+\sum_{i}\mathrm{b}_{i}n_{i}^{id}(T,\mu^{*})}.
\ee

As we consider different hardcore radii for mesons and baryons, the repulsive parameter, $b$, contributes differently in the above-mentioned thermodynamic quantities.
\subsubsection{van der Waals hadron resonance gas model}
The VDWHRG model represents a real gas system by including attractive interactions among the hadrons along with repulsive interactions due to the excluded volume of the hadrons. With these interactions, the model predicts a liquid-gas phase transition which ends at a critical point \cite{Vovchenko:2016rkn}. In the hadronic models, these interactions are assumed to exist between all pairs of baryons and all pairs of antibaryons \cite{Vovchenko:2017cbu, Vovchenko:2016rkn, Samanta:2017yhh}. As in Ref.~\cite {Singh:2025rwc, Pradhan:2022gbm}, we also consider repulsive interactions among the mesons. The baryon-antibaryon and meson-(anti)baryon VDW interactions are neglected. The VDW equation of state in the canonical ensemble can be written as \cite{Pradhan:2022gbm, Samanta:2017yhh}
\begin{equation}
\label{eq8}
    P(T,n) = \frac{nT}{1-{\rm{b}} n}- {\rm{a}} n^{2},
\end{equation}
where $a$ and $b$ (positive) are the VDW parameters that describe attractive and repulsive interactions, respectively. $P$ is the pressure of the system, whereas, $n$ = $N/V$ is the number density. The repulsive interactions are included in the first term by replacing the total volume V with the effective volume available to particles using the proper volume parameter $b = 16\pi r^{3}/3$, as in the EVHRG model. The second term takes care of the attractive interactions between particles. The model reduces to EVHRG when the attractive interaction parameter is set to zero, and further reduces to ideal HRG when both the attractive and repulsive interactions are neglected.

The VDW equation of state in the grand canonical ensemble can then be written as \cite{Samanta:2017yhh, Vovchenko:2015pya} 
\begin{equation}
\label{eq9}
    P(T,\mu) = P^{id}(T,\mu^{*}) - {\rm{a}}n^{2}(T,\mu),
\end{equation}
where the $n(T,\mu)$, the VDW particle number density can be written as
\begin{equation}
\label{eq10}
    n(T,\mu) = \frac{\sum_{i}n_{i}^{id}(T,\mu_{i}^{*})}{1+{\rm b}\sum_{i}n_{i}^{id}(T,\mu_{i}^{*})}.
\end{equation}
The $\mu^{*}$ is the modified chemical potential because of the interaction and is given by 
\begin{equation}
\label{eq11}
    \mu^{*} = \mu - {\rm b}P(T,\mu) - {\rm{ab}}n^{2}(T,\mu) + 2{\rm a}n(T,\mu).
\end{equation}

Because of the interaction among the hadrons, the contribution to total pressure in the VDWHRG model comes from the mesons, baryons, and antibaryons separately and is given by \cite{Samanta:2017yhh, Vovchenko:2016rkn} 
\begin{equation}
\label{eq14}
P(T,\mu) = P_{M}(T,\mu) + P_{B}(T,\mu) + P_{\bar{B}}(T,\mu),
\end{equation}
where the $P_{M}(T,\mu), P_{B(\bar B)}(T,\mu)$ are the pressures due to mesons and (anti)baryons, respectively and are given by,
\begin{align}
\label{eq15}
P_{M}(T,\mu) &= \sum_{i\in M}P_{i}^{id}(T,\mu^{*M}),  \\    
P_{B}(T,\mu) &= \sum_{i\in B}P_{i}^{id}(T,\mu^{*B})-{\rm a}n^{2}_{B}(T,\mu),\\
P_{\bar{B}}(T,\mu) &= \sum_{i\in \bar{B}}P_{i}^{id}(T,\mu^{*\bar{B}})-{\rm a}n^{2}_{\bar{B}}(T,\mu).
\end{align}
Here, $M$, $B$, and $\bar B$ represent mesons, baryons, and antibaryons. Because of the excluded volume correction, the modified chemical potential for mesons is given by $\mu^{*M}$, whereas, $\mu^{*B}$ and $\mu^{*\bar B}$ are the modified chemical potentials of baryons and antibaryons due to VDW interactions~\cite{Sarkar:2018mbk}. These modified chemical potentials for mesons and (anti)baryons can be obtained from Eq.~(\ref{eq2}) and Eq.~(\ref{eq11}) as, 
\begin{align}
\label{eq18}
\mu^{*M} &= -{\rm b}_MP_{M}(T,\mu),\\
\mu^{*B(\bar B)} &= \mu^{B(\bar B)}-{\rm b}_B P_{B(\bar B)}(T,\mu)-{\rm{ab}}_B n^{2}_{B(\bar B)}+2{\rm a}n_{B(\bar B)},
\end{align}
where $n_{M}$, $n_{B}$ and $n_{\bar B}$ are the modified number densities of mesons, baryons and antibaryons, respectively, which are given by
\begin{align}
\label{eq20}
    n_{M}(T,\mu) &= \frac{\sum_{i\in M}n_{i}^{id}(T,\mu_{i}^{*M})}{1+{\rm b}_M\sum_{i\in M}n_{i}^{id}(T,\mu_{i}^{*M})},\\
    n_{B(\bar B)}(T,\mu) &= \frac{\sum_{i\in B(\bar B)}n_{i}^{id}(T,\mu_{i}^{*B(\bar B)})}{1+{\rm b}_B\sum_{i\in B(\bar B)}n_{i}^{id}(T,\mu_{i}^{*B(\bar B)})}.
\end{align}

The energy density for mesons and (anti)baryons can be obtained as,
\begin{align}
\varepsilon(T,\mu)_M &= \frac{\sum_{i\in M}\varepsilon_{i}^{id}(T,\mu_{i}^{*})}{1+{\rm b}_M\sum_{i\in M}n_{i}^{id}(T,\mu_{i}^{*})}.\nn\\
\varepsilon(T,\mu)_{B(\bar B)} &= \frac{\sum_{i\in B(\bar B)}\varepsilon_{i}^{id}(T,\mu_{i}^{*})}{1+{\rm b}_B\sum_{i\in B(\bar B)}n_{i}^{id}(T,\mu_{i}^{*})} - {\rm a}n^{2}_{B(\bar B)}(T,\mu).
\end{align}\label{eq13}

\subsubsection{Repulsive mean-field hadron resonance gas model}
 In the RMFHRG model, the short-range repulsive interactions among the hadrons are taken into account in the mean field approach by considering a shift in the single-particle energies. This shift is proportional to the total number density of the hadrons. We use the repulsive mean field approach as in Ref.~\cite{Kapusta:1982qd, Olive:1980dy, Kadam:2019peo, Huovinen:2017ogf} to include the repulsive interaction. The shift in the single-particle energy because of the interaction is given by
\be
\tilde \omega_{i}=\sqrt{k_{i}^2+m_{i}^2}+U(n)=\omega_{i}+U(n).
\label{dispersion}
\ee
 Here, $U$ is the potential energy, which represents the repulsive interaction between hadrons and depends on the total hadron density $n$. For a given hadron potential $V({\bf{r}})$, the potential energy can be expressed as $U(n)=Kn$. The phenomenological parameter $K$ is determined by integrating the potential $V(\bf r)$ over the spatial volume \cite{Kapusta:1982qd, Olive:1980dy}.

In this work, we assign different repulsive interaction parameters for baryons and mesons. We denote the mean-field parameter 
for baryons ($B$) and antibaryons ($\bar{B}$) by $K_B$, while for mesons we denote it by  $K_M$. Thus,  for (anti)baryons
\be
U(n_{B\{\bar{B}\}})=K_Bn_{B\{\bar{B}\}},
\label{potenbar}
\ee
and for mesons
\be
U(n_M)=K_Mn_M.
\label{potenmes}
\ee
The repulsion parameter $K_B$ = 0.450 GeV fm$^3$ is taken to be the same for all (anti)baryons and $K_M$ = 0.050 GeV fm$^3$ for all mesons as in Ref.~\cite{Huovinen:2017ogf}.
The total hadron number density is then given by
\be
n(T,\mu)=\sum_{i}n_{i}=n_B+n_{\bar{B}}+n_M,
\ee
where $n_{i}$ is the number density of $i{\rm th}$ hadronic species. Note that $n_B$, $n_{\bar{B}}$, and $n_M$ are total baryon,
 antibaryon and meson number densities, respectively. Explicitly, for (anti)baryons,
\be
n_{B\{\bar{B}\}}=\sum_{i\in B\{\bar{B}\}}\frac{g_{i}}{2\pi^2}\int_{0}^{\infty}\:\frac{k^{2}_{i}dk_{i}}{e^{\big(\frac{\omega_{i}-\mu_{\text{eff}}^i}{T}\big)}+1}.
\label{numdenbaryon}
\ee
Here, $\mu_{\text{eff}}^i=\beta_i\mu_{B}-U(n_{B\{\bar{B}\}})$
is the (anti)baryon effective chemical potential, with $\beta_i$ being the baryonic number of $i{\rm th}$ baryon and $\mu_B$, the baryon chemical potential. For mesons (since $\mu_B$ = 0), the effective chemical potential becomes
\begin{equation}
    (\mu_{\text{eff}}^i)_{M} = -U(n_M) = -K_M n_M.
\end{equation}
Hence, the number density for mesons can be written as
\be
n_{M}=\sum_{i\in M}\frac{g_{i}}{2\pi^2}\int_{0}^{\infty}\:\frac{k^{2}_{i}dk_{i}}{e^{\big(\frac{\omega_i+K_Mn_M}{T}\big)}-1},
\label{numdenmeson}
\ee
where the sum is over all the mesons.

One can write the total (anti)baryon energy density  as
\begin{align}
\varepsilon_{B\{\bar{B}\}}=&\sum_{i\in B\{\bar{B}\}}\frac{g_{i}}{2\pi^2}\int_{0}^{\infty}\:\frac{k^{2}_{i}dk_{i}\tilde \omega_{i}}{e^{\big(\frac{\omega_{i}-\mu^i_{\text{eff}}}{T}\big)}+1}~ \nn\\
&+\phi_{B\{\bar{B}\}}(n_{B\{\bar{B}\}}),
\label{endenbaryon}
\end{align}
and for mesons
\be
\varepsilon_{M} = \sum_{i\in{M}}\frac{g_{i}}{2\pi^2}\int_{0}^{\infty}\:\frac{k^{2}_{i}dk_{i}\tilde \omega_{i}}{e^{\big(\frac{ \omega_{i}+K_Mn_M}{T}\big)}-1}+\phi_{M}(n_M),
\label{endenmeson}
\ee
where $\phi(n)$ represents the energy density correction to avoid double-counting the potential. After doing the derivative of baryon energy density with respect to baryon net number density and using Eq. (\ref{potenbar}), we get

\be
\frac{\partial \phi_{B\{\bar{B}\}}}{\partial n_{B\{\bar{B}\}}}=-K_Bn_{B\{\bar{B}\}},
\ee

and hence

\be
\phi_B(n_{B\{\bar{B}\}})=-\frac{1}{2}K_Bn_{B\{\bar{B}\}}^2.
\ee

Also, for mesons, one can obtain in a similar way
\be
\phi_M(n_M)=-\frac{1}{2}K_Mn_M^2.
\ee

The pressure of the gas can now be readily obtained. For baryons
\begin{align}
P_{B\{\bar{B}\}}(T,\mu)=&T\sum_{i\in B\{\bar{B}\}}\frac{g_{i}}{2\pi^2}\int_{0}^{\infty}kd^{2}_{i}k_{i}\nn\\ 
&\text{ln}\bigg[1 + e^{-\big(\frac{\omega_i-\mu^i_{\text{eff}}}{T}\big)}\bigg]
-\phi_{B\{\bar{B}\}}(n_{B\{\bar{B}\}}),
\end{align}
and for mesons
\begin{align}
    P_M(T)=&-T\sum_{i\in M}\frac{g_{i}}{2\pi^2}\int_{0}^{\infty}k^{2}_{i}dk_{i}\nn\\
    &\text{ln}\bigg[1- e^{-\big(\frac{ \omega_{i}+K_Mn_M}{T}\big)}\bigg]-\phi_M(n_M). 
\end{align}

\subsection{Leading-order thermoelectric coefficients}
\label{formalism2}
To determine the leading-order thermoelectric coefficients in hot hadron gas, we begin with the relativistic Boltzmann transport equation under the relaxation time approximation. It provides a kinetic theory framework that tracks the phase-space distribution of particles, allowing one to model time-dependent processes like thermalization and transport. The presence of a magnetic field modifies the phase-space dynamics of charged particles through the Lorentz force, which introduces anisotropy in both electric and heat transport.  Within RTA, the RBTE can be considered as a linear expansion of the single-particle total distribution function ($f_i$) around the single-particle equilibrium distribution function ($f_i^{0}$). Here, the total distribution function can be written as $f_i=f^0_i+\delta f_i$, where $\delta f_i$ is a non-equilibrium correction that represents the deviation from equilibrium. In the HRG model, each hadron species contributes to the total distribution function. For hadron species $i$ with energy $\omega_i=\sqrt{\vec{k_i}^2+m_i^2}$, baryon number $\beta_i$, and the baryon chemical potential $\mu_B$, the total single-particle distribution function is
\begin{align}\label{Dis-f}
f^0_i= \frac{1}{e^{\frac{\omega_i-\beta_i\mu_B}{T}}\pm 1}~,
\end{align}
where the $\pm$ stands for fermion and boson, respectively. In local thermodynamic equilibrium, the spatial dependence of the distribution function appears due to spatial gradients of temperature and baryon chemical potential.
The RBTE for a single hadron species under RTA in the presence of an electromagnetic field can be expressed as~\cite{Das:2020beh, Singh:2023pwf}
\begin{align}
k^\mu \partial_\mu f_i + q_i F^{\mu\nu} k_\nu \frac{\partial f_i}{\partial k^\mu} = -\frac{k^\mu u_\mu}{\tau_R^i} \delta f_i,
\label{equnew21}
\end{align}
where \(k^\mu = (\omega_i, \vec{k}_i)\) is the particle four-momentum, \(u^\mu\) is the fluid four-velocity, \(F^{\mu\nu}\) is the electromagnetic field tensor, and \(\tau_R^i\) is the relaxation time for species \(i\).

To solve the above equation, we take an ansatz to express the deviation of the distribution function from the equilibrium, such that it should reflect the driving forces. The general form of ansatz is~\cite{Singh:2023pwf, Singh:2023ues, Singh:2024emy,Gavin:1985ph}
\begin{align}
\delta f_i =  \frac{\partial f_i^0}{\partial (k^\mu u_\mu)} k^\mu \Omega_\mu,
\label{equnew22T}
\end{align}
where the vector \(\Omega^\mu\) reflects the response of the distribution function to external thermodynamic forces such as temperature gradients, electric fields, and magnetic fields. In the fluid rest frame (\(u^\mu = (1, \vec{0})\)), \(\Omega^\mu\) can be decomposed as
\begin{align}
\Omega^\mu &= \alpha_1 E^\mu + \alpha_2 B^\mu + \alpha_3 \epsilon^{\mu\nu\rho\sigma} u_\nu E_\rho B_\sigma + \alpha_4 \nabla^\mu T\nn\\
&+ \alpha_5 \epsilon^{\mu\nu\rho\sigma} u_\nu \nabla_\rho T B_\sigma + \alpha_6 \epsilon^{\mu\nu\rho\sigma} u_\nu \nabla_\rho T E_\sigma,
\label{equnew23T}
\end{align}
where the electric and magnetic fields in the fluid rest frame are
\[
E^\mu = F^{\mu\nu} u_\nu, \quad B^\mu = \frac{1}{2} \epsilon^{\mu\nu\rho\sigma} u_\nu F_{\rho\sigma},
\]
and \(\nabla^\mu = \Delta^{\mu\nu} \partial_\nu\) is the spatial gradient orthogonal to \(u^\mu\), with \(\Delta^{\mu\nu} = g^{\mu\nu} - u^\mu u^\nu\). Here, $\epsilon^{\mu\nu\rho\sigma}$ denotes the Levi-Civita symbol, which is used to represent antisymmetric tensor components. 

 Using Eq.\eqref{equnew22T} into RBTE as given in  Eq.\eqref{equnew21}  and projecting onto spatial components in the local rest frame yields~\cite{Singh:2025rwc},
\begin{align}
&\vec{v}_i \cdot\bigg[- \frac{\partial f^{0}_{i}}{\partial\omega_i}\bigg(\frac{\omega_i-\beta_ih}{T}\bigg)\vec{\nabla}T\bigg]+q_i(\vec{E} \cdot \vec{v}_i)\frac{\partial f^{0}_{i}}{\partial \omega_i}\nonumber\\
&-q_i \vec{v}_i \cdot(\vec{\Omega}\times\vec{B})\frac{\partial f^{0}_{i}}{\partial \omega_i} = -\frac{\omega_i}{\tau^{i}_R}(\vec{v}_i \cdot\vec{\Omega})\frac{\partial f^{0}_{i}}{\partial \omega_i}.
\label{equnew24T}
\end{align}
Where \(\vec{v}_i = \vec{k}_i / \omega_i\) is the relativistic particle velocity and $h = \frac{\epsilon + P}{n}$ is the enthalpy per baryon, $\epsilon, P$, and $n$ are total energy density, total pressure, and net baryon density of the system, respectively. Here, the enthalpy per baryon is induced in the picture following the Gibbs-Duhem relation. Using the Gibbs--Duhem relation, the gradients of the inverse temperature and the baryon chemical potential are combined, allowing the gradient of the equilibrium distribution function to be rewritten solely in terms of the temperature gradient, with the enthalpy per baryon naturally emerging as the relevant thermodynamic quantity~\cite{Singh:2023ues, Singh:2024emy, Das:2020beh, Singh:2025geq}.

By substituting $\vec{\Omega}$ from Eq.\eqref{equnew23T} into Eq.\eqref{equnew24T}, we can expand it analogously to the nonrelativistic case as,
\begin{align}
 &q_i(\vec{E} \cdot \vec{v}_i)-\alpha_1 q_i\vec{v}_i \cdot(\vec{E}\times\vec{B})-\alpha_3 q_i  (\vec{E} \cdot \vec{B})(\vec{v}_i \cdot\vec{B})+\alpha_3 \nonumber\\
 &q_i (\vec{v}_i \cdot\vec{E})-\alpha_4 q_i  \vec{v}_i \cdot(\vec{\nabla}T\times \vec{B})-\alpha_5q_i(\vec{\nabla}T \cdot \vec{B})(\vec{v}_i \cdot\vec{B})+\alpha_5\nonumber\\
 &q_i(\vec{v}_i \cdot\vec{\nabla}T)-\alpha_6q_i(\vec{\nabla}T \cdot \vec{B})(\vec{v}_i \cdot\vec{E})+\alpha_6q_i(\vec{E} \cdot \vec{B})(\vec{v}_i \cdot\vec{\nabla}T)\nonumber\\
 &-\bigg(\frac{\omega_i-\beta_ih}{T}\bigg)(\vec{v}_i \cdot\vec{\nabla} T)
  = -\frac{\omega_i}{\tau^{i}_R}\bigg[\alpha_1(\vec{v}_i \cdot\vec{E})+\alpha_2(\vec{v}_i \cdot\vec{B})\nonumber\\
 & +\alpha_3 \vec{v}_i \cdot(\vec{E}\times\vec{B})+\alpha_4(\vec{v}_i \cdot\vec{\nabla}T)+\alpha_5\vec{v}_i \cdot(\vec{\nabla}T\times \vec{B})+\nonumber\\&\alpha_6\vec{v}_i \cdot(\vec{\nabla}T\times\vec{E})\bigg].
 \label{equnew25}
\end{align}
Comparing the coefficients of same tensor structures on both sides of Eq.\eqref{equnew25} we get,
\begin{align}
 & \alpha_1 = -\frac{\tau^{i}_R}{\omega_i}(q_iE+\alpha_3 q_iB)\label{equnew27},\\
 & \alpha_2 = \alpha_3 \tau^{i}_R \Omega_{c_i}(\vec{E} \cdot \vec{B})+b\tau^{i}_R\Omega_{c_i}(\vec{\nabla}T \cdot \vec{B}),\label{equnew29}\\
 & \alpha_3 = \alpha_1 \tau^{i}_R \Omega_{c_i},\label{equnew28}\\
 & \alpha_4 =\frac{\tau^{i}_R}{\omega_i}\bigg(\frac{\omega_i-\beta_ih}{T}\bigg)-\alpha_5 \tau^{i}_R\Omega_{c_i},\\
 & \alpha_5= \Omega_{c_i}\tau^{i}_R\alpha_4,\label{equnew30}\\
 & \alpha_6=0
 \label{equnew31}.
\end{align}
where $\Omega_{c_i}=\frac{q_iB}{\omega_i}$ represents the cyclotron frequency of the particle with electric charge $q_i$. Using Eqs.\eqref{equnew27} to \eqref{equnew31}, we get
\begin{align}
 & \alpha_1 = -\frac{(q_iE)(\tau^{i}_R/\omega_i)}{1+(\Omega_{c_i}\tau^{i}_R)^2},\label{equnew33}\\
 & \alpha_4 = \frac{\tau^{i}_R}{\omega_i}\bigg(\frac{\omega_i- \beta_ih}{T}\bigg)\frac{1}{1+(\Omega_{c_i}\tau^{i}_R)^2}\label{equnew34}.
\end{align}
Using $\alpha_1$ and $\alpha_4$ as given in Eq.\eqref{equnew33} and Eq.\eqref{equnew34}, we can write
deviation from the equilibrium distribution function as,
\begin{align}
 \delta f_i &= \frac{\tau^{i}_R}{1+(\Omega_{c_i}\tau^{i}_R)^2} \bigg[q_i\bigg\{(\vec{v}_i \cdot\vec{E})+(\Omega_{c_i}\tau^{i}_R)\vec{v}_i \cdot(\vec{E}\times\vec{B})+\nonumber\\
& (\Omega_{c_i}\tau^{i}_R)^2(\vec{E} \cdot \vec{B})(\vec{v}_i \cdot\vec{B})\bigg\}
  -\bigg(\frac{\omega_i- \beta_ih}{T}\bigg)\bigg\{(\vec{v}_i \cdot\vec{\nabla}T)+\nonumber\\
  &(\Omega_{c_i}\tau^{i}_R)\vec{v}_i \cdot(\vec{\nabla} T\times B)+(\Omega_{c_i}\tau^{i}_R)^2(\vec{\nabla}T \cdot \vec{B})(\vec{v}_i \cdot\vec{B})\bigg\}\bigg](-\frac{\partial f^{0}_{i}}{\partial\omega_i}).
 \label{equnew35}
 \end{align}
Now, we can express the electrical current and the heat current using $\delta f_i$ as given in Eq.\eqref{equnew35} as,
 \begin{align}
  j^l & =\sum_i g_i \int \frac{d^3k_i}{(2\pi)^3}q_i v_i^{l}\delta f_i\nonumber\\
  & = \sum_i \frac{g_iq_i}{3}\int\frac{d^3k_i}{(2\pi)^3}\frac{v_i^2 \tau^{i}_R}{1+(\Omega_{c_i}\tau^{i}_R)^2}\bigg[ q_i\delta^{lj}E^j+q_i (\Omega_{c_i}\tau^{i}_R)\nonumber\\
  &\epsilon^{ljk}h^kE^j+q_i(\Omega_{c_i}\tau^{i}_R)^2h^lh^jE^j
   -\bigg(\frac{\omega_i- \beta_ih}{T}\bigg)\bigg\{\delta^{lj}\frac{\partial T}{\partial x^j}\nonumber\\
  & +(\Omega_{c_i}\tau^{i}_R)\epsilon^{ljk}h^k \frac{\partial T}{\partial x^j}+(\Omega_{c_i}\tau^{i}_R)^2h^lh^j\frac{\partial T}{\partial x^j}\bigg\}\bigg](-\frac{\partial f_i^{0}}{\partial \omega_i}),
  \label{equnew36}
 \end{align}
and,
\begin{align}
 {I}^l &  = \sum_i g_i \int \frac{d^3k_i}{(2\pi)^3}v_i^l\bigg(\omega_i-\beta_ih\bigg)\delta f_i\nonumber\\
 & = \sum_i \frac{g_i}{3}\int\frac{d^3k_i}{(2\pi)^3}\frac{v_i^2\tau^{i}_R}{1+(\Omega_{c_i}\tau^{i}_R)^2}\bigg(\omega_i-\beta_ih\bigg)\bigg[ q_i\delta^{lj}E^j \nonumber\\
 &+q_i (\Omega_{c_i}\tau^{i}_R)\epsilon^{ljk}h^kE^j+q_i(\Omega_{c_i}\tau^{i}_R)^2h^lh^jE^j
  -\bigg(\frac{\omega_i- \beta_ih}{T}\bigg)\nonumber\\
  &\bigg\{\delta^{lj}\frac{\partial T}{\partial x^j}+(\Omega_{c_i}\tau^{i}_R)\epsilon^{ljk}h^k \frac{\partial T}{\partial x^j}+(\Omega_{c_i}\tau^{i}_R)^2h^lh^j\frac{\partial T}{\partial x^j}\bigg\}\bigg]\nonumber\\
  &~~~~~~~~~~~~~~~~~~~~~~~~~~~~~~~~~~~~~~~~~~~~~~~~~~~~~~~~~(-\frac{\partial f_i^{0}}{\partial \omega_i}).
  \label{equnew37}
\end{align}
Here, to simplify further calculations,  we can choose the magnetic field along the $z$ direction. The direction of the electric field and the temperature gradient are perpendicular to the $z$ axis, i.e., it is in the $x-y$ plane. Under these conditions, the components of the electric current in the $x-y$ plane are given as,
\begin{align}
 j_x = &  \sum_i\frac{g_iq_i}{3}\int\frac{d^3k_i}{(2\pi)^3}\frac{v_i^2q_i\tau^{i}_R}{1+(\Omega_{c_i}\tau^{i}_R)^2}\bigg[E_x+(\Omega_{c_i}\tau^{i}_R)E_y\bigg]\nonumber\\
 &(-)\frac{\partial f_i^{0}}{\partial \omega_i}
  -\sum_i\frac{g_i q_i}{3T}\int\frac{d^3k_i}{(2\pi)^3}\frac{v_i^2\tau^{i}_R\left(\omega_i-\beta_ih\right)}{1+(\Omega_{c_i}\tau)^2}\nonumber\\
 & \bigg[\frac{dT}{dx}+(\Omega_{c_i}\tau^{i}_R)\frac{dT}{dy}\bigg]
 (-\frac{\partial f_i^{0}}{\partial \omega_i}),
 \label{equnew38}
\end{align}
and,
\begin{align}
 j_y = &  \sum_i\frac{g_iq_i}{3}\int\frac{d^3k_i}{(2\pi)^3}\frac{v_i^2q_i\tau^{i}_R}{1+(\Omega_{c_i}\tau^{i}_R)^2}\bigg[E_y-(\Omega_{c_i}\tau^{i}_R)E_x\bigg]\nonumber\\
 &(-)\frac{\partial f_i^{0}}{\partial \omega_i}
  -\sum_i\frac{g_i q_i}{3T}\int\frac{d^3k_i}{(2\pi)^3}\frac{v_i^2\tau^{i}_R\left(\omega_i-\beta_ih\right)}{1+(\Omega_{c_i}\tau)^2}\nonumber\\
  &\bigg[\frac{dT}{dy}-(\Omega_{c_i}\tau^{i}_R)\frac{dT}{dx}\bigg](-\frac{\partial f_i^{0}}{\partial \omega_i}).
 \label{equnew39}
\end{align}
Eqs.\eqref{equnew38} and \eqref{equnew39} can be written in a compact form by introducing the following integrals,
\begin{align}
 & H_{1_i}=\frac{g_i}{3}\int\frac{d^3k_i}{(2\pi)^3}\frac{\tau^{i}_R}{1+(\Omega_{c_i}\tau^{i}_R)^2}\left(\frac{\vec{k_i}^2}{\omega_i^2}\right)(-\frac{\partial f_i^{0}}{\partial \omega_i}),\label{equnew40}\\
& H_{2_i}=\frac{g_i}{3}\int\frac{d^3k_i}{(2\pi)^3}\frac{\tau^{i}_R(\Omega_{c_i}\tau^{i}_R)}{1+(\Omega_{c_i}\tau^{i}_R)^2}\left(\frac{\vec{k_i}^2}{\omega_i^2}\right)(-\frac{\partial f_i^{0}}{\partial \omega_i}),\label{equnew41}\\
& H_{3_i}=\frac{g_i}{3}\int\frac{d^3k_i}{(2\pi)^3}\frac{\tau^{i}_R\omega_i}{1+(\Omega_{c_i}\tau^{i}_R)^2}\left(\frac{\vec{k_i}^2}{\omega_i^2}\right)(-\frac{\partial f_i^{0}}{\partial \omega_i}),\label{equnew42}\\
& H_{4_i}=\frac{g_i}{3}\int\frac{d^3k_i}{(2\pi)^3}\frac{\tau^{i}_R\omega_i(\Omega_{c_i}\tau^{i}_R)}{1+(\Omega_{c_i}\tau^{i}_R)^2}\left(\frac{\vec{k_i}^2}{\omega_i^2}\right)(-\frac{\partial f_i^{0}}{\partial \omega_i}).\label{equnew43}
 \end{align}
The integrals as given in Eqs.\eqref{equnew40}-\eqref{equnew43} allows us to write  Eq.\eqref{equnew38} and  Eq.\eqref{equnew39}, respectively, as 
\begin{widetext}
\begin{align}
 j_x = \sum_i q_i^2H_{1_i}E_x+\sum_i q_i^2H_{2_i}E_y -\frac{1}{T}\sum_{a}q_i\bigg(H_{3_i}-\beta_ihH_{1_i}\bigg)\frac{d T}{dx}-\frac{1}{T}\sum_{a}q_i\bigg(H_{4_i}-\beta_ihH_{2_i}\bigg)\frac{d T}{dy},
 \label{equnew44}
\end{align}
and,
\begin{align}
 j_y = \sum_i q_i^2H_{1_i}E_y-\sum_i q_i^2H_{2_i}E_x -\frac{1}{T}\sum_{a}q_i\bigg(H_{3_i}-\beta_ihH_{1_i}\bigg)\frac{d T}{dy}+\frac{1}{T}\sum_{a}q_i\bigg(H_{4_i}-\beta_ihH_{2_i}\bigg)\frac{d T}{dx}.
 \label{equnew45}
\end{align}
Here, in the presence of a magnetic field, the magneto-Seebeck coefficient ($S_B$) can be determined by setting $j_x = j_y=0$ so that the electric field becomes proportional to the temperature gradient. For $j_x=0$ and $j_y=0$ we can solve Eqs.\eqref{equnew44} and \eqref{equnew45} to get $E_x$ and $E_y$ in terms of temperature gradients $\frac{dT}{dx}$ and  $\frac{dT}{dy}$ as,
\begin{align}
 E_x &  = \frac{\sum_i q_i^2H_{1_i}\sum_iq_i(H_{3_i}-\beta_ihH_{1_i})+\sum_i q_i^2H_{2_i}\sum_iq_i(H_{4_i}-\beta_ihH_{2_i})}{T\bigg[\bigg(\sum_i q_i^2H_{1_i}\bigg)^2+\bigg(\sum_i q_i^2H_{2_i}\bigg)^2\bigg]}\frac{dT}{dx}\nonumber\\
 & +\frac{\sum_i q_i^2H_{1_i}\sum_iq_i(H_{4_i}-\beta_ihH_{2_i})-\sum_i q_i^2H_{2_i}\sum_iq_i(H_{3_i}-\beta_ihH_{1_i})}{T\bigg[\bigg(\sum_i q_i^2H_{1_i}\bigg)^2+\bigg(\sum_i q_i^2H_{2_i}\bigg)^2\bigg]}\frac{dT}{dy},
 \label{equnew46}
\end{align}
and,
\begin{align}
 E_y &  = \frac{\sum_i q_i^2H_{2_i}\sum_iq_i(H_{3_i}-\beta_ihH_{1_i})-\sum_i q_i^2H_{1_i}\sum_iq_i(H_{4_i}-\beta_ihH_{2_i})}{T\bigg[\bigg(\sum_i q_i^2H_{1_i}\bigg)^2+\bigg(\sum_i q_i^2H_{2_i}\bigg)^2\bigg]}\frac{dT}{dx}\nonumber\\
 & +\frac{\sum_i q_i^2H_{1_i}\sum_iq_i(H_{3_i}-\beta_ihH_{1_i})+\sum_i q_i^2H_{2_i}\sum_iq_i(H_{4_i}-\beta_ihH_{2_i})}{T\bigg[\bigg(\sum_i q_i^2H_{1_i}\bigg)^2+\bigg(\sum_i q_i^2H_{2_i}\bigg)^2\bigg]}\frac{dT}{dy}.
 \label{equnew47}
\end{align}
\end{widetext}
Eqs.\eqref{equnew46} and \eqref{equnew47} can be written in a compact form in the following way,
\begin{align}
 \begin{pmatrix}
E_x \\
\\
E_y 
\end{pmatrix}= \begin{pmatrix}
S_B & NB \\
\\
-NB & S_B 
\end{pmatrix}\begin{pmatrix}
\frac{dT}{dx} \\
\\
\frac{dT}{dy} 
\end{pmatrix},
\end{align}
here one can identify the magneto-Seebeck coefficient as, 
\begin{widetext}
\begin{align}
 S_B & = \frac{\sum_i q_i^2H_{1_i}\sum_iq_i(H_{3_i}-\beta_ihH_{1_i})+\sum_i q_i^2H_{2_i}\sum_iq_i(H_{4_i}-\beta_ihH_{2_i})}{T\bigg[\bigg(\sum_i q_i^2H_{1_i}\bigg)^2+\bigg(\sum_i q_i^2H_{2_i}\bigg)^2\bigg]}
  = \frac{(\sigma_{el}/T)(\mathcal{I}_{31}/T^2)+(\sigma_{H}/T)(\mathcal{I}_{42}/T^2)}{(\sigma_{el}/T)^2+(\sigma_{H}/T)^2},
 \label{equnew49}
\end{align}
and the normalized Nernst coefficient ($NB$) is given as,
\begin{align}
 NB & = \frac{\sum_i q_i^2H_{1_i}\sum_iq_i(H_{4_i}-\beta_ihH_{2_i})-\sum_i q_i^2H_{2_i}\sum_iq_i(H_{3_i}-\beta_ihH_{1_i})}{T\bigg[\bigg(\sum_i q_i^2H_{1_i}\bigg)^2+\bigg(\sum_i q_i^2H_{2_i}\bigg)^2\bigg]}
  = \frac{(\sigma_{el}/T)(\mathcal{I}_{42}/T^2)-(\sigma_{H}/T)(\mathcal{I}_{31}/T^2)}{(\sigma_{el}/T)^2+(\sigma_{H}/T)^2}.
 \label{equnew50}
 \end{align}
 \end{widetext}
Here, we have identified the ohmic-like component of electrical conductivity
in the presence of a magnetic field and the Hall-like component of electrical conductivity as $\sigma_{el}=\sum_iq_i^2H_{1_i}$ and $\sigma_{H}=\sum_iq_i^2H_{2_i}$ respectively \cite{Singh:2024emy}. The integrals $\mathcal{I}_{31}$ and $\mathcal{I}_{42}$ in Eqs.\eqref{equnew49} and \eqref{equnew50} are defined as  $\mathcal{I}_{31} = \sum_iq_i(H_{3_i}-\beta_ihH_{1_i})$ and $\mathcal{I}_{42} = \sum_iq_i(H_{4_i}-\beta_ihH_{2_i})$.
Note that in the absence of a magnetic field, integrals $H_{2_i}$ and $H_{4_i}$ are identically zero. Hence, the normalized Nernst coefficient vanishes in the absence of a magnetic field, and the magneto-Seebeck coefficient reduces to the Seebeck coefficient in the absence of a magnetic field~\cite{Singh:2025rwc}. 
The coefficients $S_B$ and $NB$ have temperature dependence as shown in Eqs.~\ref{equnew49} and \ref{equnew50}, hence the higher-order thermoelectric coefficients are introduced for the medium.  
\subsection{Higher-order thermoelectric coefficients}
Higher-order thermoelectric coefficients differ from their leading-order counterparts in behavior. While leading-order effects such as the Seebeck or Peltier effects arise from linear responses to small gradients in temperature or electric field, higher-order phenomena manifest when systems are subjected to non-linear responses. The temperature dependence of the leading-order thermoelectric coefficient, such as the Seebeck coefficient, further gives rise to the higher-order thermoelectric coefficients.  The presence of the leading-order thermoelectric coefficient in the conducting medium modifies the charge current and heat current of the medium as~\cite{Das:2020beh, Singh:2025rwc}
\begin{align}\label{ji}
    \vec j &= \sigma_{el}\vec{E} - \sigma_{el}S\vec{\nabla}T,\nn\\
    \vec I &= T\sigma_{el}S\vec{E} - \kappa_0\vec{\nabla}T,
\end{align}
where $\kappa_0$ is the thermal conductivity of the medium and is expressed as~\cite{Singh:2025rwc}
\begin{align}
 \kappa_0=\sum_i\frac{g_i}{3T^2}\int\frac{d^3k_i}{(2\pi)^3}\tau^{i}_R\left(\frac{\vec{k_i}}{\omega_i}\right)^2\left(\omega_i-\beta_ih\right)^2f^{0}_{i}(1\mp f^{0}_{i}).
 \label{equnew15}
\end{align}
Using Eq.\eqref{ji}, we can express the heat current $\vec{{I}}$ in terms  of electric current $\vec{j}$ in the following way,
\begin{equation}
 \vec{{I}}=TS\vec{j}-\left(\kappa_0-T\sigma_{el}S^2\right)\vec{\nabla}T.
 \label{equnew16}
\end{equation}
Here, $\kappa^{'} = \kappa_0-T\sigma_{el}S^2$ is defined as the modified thermal conductivity of the medium. The Eq.\eqref {ji} can be defined in matrix form as
\begin{align}
 \begin{pmatrix}
\vec{j} \\
\\
\vec{I} 
\end{pmatrix}= \begin{pmatrix}
\sigma_{el} &  \sigma_{el}ST\\
\\
\sigma_{el}ST & \kappa_0T 
\end{pmatrix}\begin{pmatrix}
\vec{E} \\
\\
-\vec\nabla T/T 
\end{pmatrix}.
\end{align}
The above matrix can be rearranged and expressed as follows
\begin{align}
 \begin{pmatrix}
\vec{E} \\
\\
\vec{I} 
\end{pmatrix}= \begin{pmatrix}
\rho_{el} &  S\\
\\
ST & -\kappa^{'}
\end{pmatrix}\begin{pmatrix}
\vec{j} \\
\\
-\vec\nabla T/T 
\end{pmatrix},
\end{align}
where resistivity of the medium is $\rho_{el} = 1/\sigma_{el}$. When a magnetic field is introduced in the z direction, electric field and temperature gradients in x-y plane, the above matrix in x-y plane can be expressed as
\begin{align}
 \begin{pmatrix}
{E_x} \\
\\
{E_y} \\
\\
I_x\\ 
\\
I_y
\end{pmatrix}= \begin{pmatrix}
\rho_{el} & \rho_{H} &  S_B &  NB\\
\\
-\rho_{H} & \rho_{el} & -NB & S_B\\
\\
S_BT & NBT & -\kappa^{'}_{el} & -\kappa^{'}_{H}\\
\\
- NBT &  S_BT & \kappa^{'}_{H} & -\kappa^{'}_{el}
\end{pmatrix}\begin{pmatrix}
j_x \\
\\
j_y \\
\\
\frac{dT}{dx}\\
\\
\frac{dT}{dy}\\
\end{pmatrix},
\end{align}
The $\kappa_{el}$ and $\kappa_H$ are Ohmic-like and Hall-like components of thermal conductivity of the medium in the presence of a magnetic field~\cite{Singh:2023ues}, such that $\kappa^{'}_{el} = \kappa_{el}-T\sigma_{el}S_B^2$ and $\kappa^{'}_{H} = \kappa_{H}-T\sigma_{H}(NB)^2$. Also the $\rho_{H} = 1/\sigma_{H}$.

For any conducting medium having non-zero current density and temperature gradients along the x direction and an applied magnetic field along the z direction, the heat production rate per unit volume ($\dot Q$) can be expressed as~\cite{takahagi}
\begin{align} \label{heat}
    \dot Q &= E_xj_x - \frac{dI_x}{dx}\nn\\
    &= (\rho_{el}j_x + S_B\frac{dT}{dx})j_x - \frac{dT}{dx}(S_BTj_x - \kappa^{'}_{el}\frac{dT}{dx}) \nn\\
    &= \rho_{el}j_x^2 + \frac{dT}{dx}(\kappa^{'}_{el}\frac{dT}{dx}) - (T\frac{dS_B}{dT})j_x\frac{dT}{dx}.
\end{align}
The last term in the above equation is the heat generated by the combined action of $j_x$ and $\frac{dT}{dx}$. The heat production rate per unit volume due to the Thomson effect in the presence of a magnetic field is given by~\cite{ioffe1957physics}
\begin{align}\label{heat_th}
   \dot Q = -Th_B ~(\vec{j}\cdot\vec\nabla T)
\end{align}
Finally, from Eqs.~\ref{heat} and \ref{heat_th} we can define the magneto-Thomson coefficient ($Th_B$) as
\begin{align}\label{Mag-Thomson}
    Th_B = T\frac{dS_B}{dT}.
\end{align}
In the absence of a magnetic field, the coefficient $Th_B$ reduces to the Thomson coefficient $Th$~\cite{Singh:2025rwc, Singh:2025geq}.

The transverse Thomson coefficient ($Th_N$) is introduced in the medium because of the presence of a finite value of the normalized Nernst coefficient. In general, the coefficient $Th_N$ is expected to occur when a charge current, temperature gradient, and magnetic field are oriented perpendicular to each other in any conducting medium. We consider the heat production rate per unit volume when the current density is along the x direction and temperature gradients are along the y direction, whereas a magnetic field is directed along the z axis. Hence, the heat production rate per unit volume is 
\begin{align} \label{heat1}
    \dot Q &= E_xj_x - \frac{dI_y}{dy}\nn\\
    &= (\rho_{el}j_x + NB\frac{dT}{dy})j_x - \frac{dT}{dy}(-NBTj_x - \kappa^{'}_{el}\frac{dT}{dy}) \nn\\
    &= \rho_{el}j_x^2 + \frac{dT}{dy}(\kappa^{'}_{el}\frac{dT}{dy}) + (T\frac{d(NB)}{dT} + 2NB)j_x\frac{dT}{dy}.
\end{align}
Therefore, the transverse Thomson coefficient is defined as
\begin{align}\label{Trans-Thomson}
    Th_N = T\frac{d(NB)}{dT} + 2NB.
\end{align}
Unlike the $Th_B$, the coefficient $Th_N$ not only depends on the temperature derivative of the leading thermoelectric coefficient (for this case $NB$) but also on its magnitude~\cite{takahagi}.
In the absence of a magnetic field, the coefficient $Th_N$ vanishes because of the vanishing $NB$.

In the above calculation, for the sake of simplicity, we take the thermal-averaged relaxation time after integrating the energy-dependent relaxation time over the equilibrium distribution function. The thermal averaged relaxation time ($\tau_{R}^i$) for the {\it i}th hadron species can be expressed in terms of scattering cross-section as~\cite{Das:2020beh},
\begin{align}
    {\tau_{R}^i}^{-1} = \sum_j n_j \langle\sigma_{ij}v_{ij}\rangle
\end{align}
where,
\begin{align}
 \langle\sigma_{ij}v_{ij}\rangle &= \frac{\sigma}{8Tm_i^2m_j^2\mathcal{K}_2(m_i/T)\mathcal{K}_2(m_j/T)}~ \times\nn\\
 &\int_{(m_i+m_j)^2}^{\infty}ds~\times \nn\\
 &\frac{[s-(m_i-m_j)^2]}{\sqrt{s}}
 \times [s-(m_i+m_j)^2]\mathcal{K}_1(\sqrt{s}/T),
 \label{equnew70}
\end{align}
Here, $\sigma = 4\pi r_h^2$ is the total scattering cross-section for the hard spheres, and it is independent of both temperature and baryon chemical potential. $\mathcal{K}_1, \mathcal{K}_2$ are modified Bessel functions of the first and second order.
\subsection{Landau quantization of charged hadrons in a magnetic field}

In the presence of an external magnetic field, say along the z-axis, the dynamics of charged hadrons undergo a fundamental change due to Landau quantization. The transverse motion is no longer continuous but restricted to discrete Landau levels, while the longitudinal motion along the field direction remains continuous. This leads to a modification of the phase–space structure, such that the standard three-momentum integration is replaced by a summation over Landau levels and spin projections combined with an integration over the longitudinal momentum,
\begin{equation}
\int \frac{d^3k_i}{(2\pi)^3} \longrightarrow 
\frac{|q_i|B}{2\pi} \sum_{l=0}^{\infty} \sum_{s_z} \int_{-\infty}^{\infty} \frac{dk_{z}}{2\pi},
\end{equation}
where $q_i$ is the electric charge of the $i$th hadron, $B$ denotes the external magnetic field strength, $l$ labels the Landau levels, and $s_z$ represents the spin projection. Here, we consider the hadrons having the spins 0, 1/2, and 1~\cite{Sahoo:2023vkw}. For the charged hadrons, we also consider in the transverse plane, $k_x^2\approx k_y^2\approx (\frac{k_x^2+k_y^2}{2})=\frac{2l|{q_i}|B}{2}$~\cite{Singh:2024emy}.

The corresponding single–particle dispersion relation also acquires a discrete structure,
\begin{equation}
\omega_{c,i}^z(k_{z_i},l,s_{zi}) = \sqrt{k_{z_i}^{2} + m_i^{2} + 2(l+1/2-s_{zi})|q_i|B},
\end{equation}
which encodes the discrete orbital quantization due to the magnetic field together with the spin–magnetic field coupling. In the strong–field limit, the system approaches the lowest Landau level (LLL) approximation, where only the $l=0$ mode survives, effectively reducing the dimensionality of the momentum space.

These modifications have several important consequences. First, the density of states of charged hadrons is reshaped, altering the particle number distribution and bulk thermodynamic quantities such as pressure and energy density. Second, since scattering probabilities depend on the available phase space, the relaxation times are significantly modified in a magnetic background. As a result, all transport coefficients acquire non-trivial corrections that encode the anisotropy introduced by the magnetic field.

Our analysis incorporates Landau quantization explicitly in the hadronic phase, thereby ensuring that both the thermodynamic inputs and the transport coefficients reflect the microscopic changes induced by magnetic fields. This provides a more realistic and consistent description of the magnetic response of hadronic matter. To estimate the higher-order thermoelectric coefficients in the presence of a magnetic field, the defined integrals in Eqs.~\eqref{equnew40}-\eqref{equnew43}, for the charged hadrons, get modified as
\begin{align}
  H_{1_{c,i}}^z&= g_i\frac{|q_i|B}{2\pi} \sum_{l=0}^{\infty} \sum_{s_z} \int_{-\infty}^{\infty} \frac{dk_{z}}{2\pi}\nn\\
 &~~~~~~~~~~~~~~~~~~~~~\frac{\tau^{i}_R}{1+(\Omega_{c_i}\tau^{i}_R)^2}\left(\frac{{l|{q_i}|B}}{\omega_i^2}\right)
(-\frac{\partial f_i^{0}}{\partial \omega_i}),\label{equnew82}\\
 H_{2_{c,i}}^z&=g_i\frac{|q_i|B}{2\pi} \sum_{l=0}^{\infty} \sum_{s_z} \int_{-\infty}^{\infty} \frac{dk_{z}}{2\pi}\nn\\
 &~~~~~~~~~~~~~~~~~~~~~~\frac{\tau^{i}_R(\Omega_{c_i}\tau^{i}_R)}{1+(\Omega_{c_i}\tau^{i}_R)^2}\left(\frac{{l|{q_i}|B}}{\omega_i^2}\right)
(-\frac{\partial f_i^{0}}{\partial \omega_i}),\label{equnew83}\\
 H_{3_{c,i}}^z&=g_i\frac{|q_i|B}{2\pi} \sum_{l=0}^{\infty} \sum_{s_z} \int_{-\infty}^{\infty} \frac{dk_{z}}{2\pi}\nn\\
 &~~~~~~~~~~~~~~~~~~~~~~\frac{\tau^{i}_R\omega_i}{1+(\Omega_{c_i}\tau^{i}_R)^2}\left(\frac{{l|{q_i}|B}}{\omega_i^2}\right)
(-\frac{\partial f_i^{0}}{\partial \omega_i}),\label{equnew84}\\
 H_{4_{c,i}}^z&=g_i\frac{|q_i|B}{2\pi} \sum_{l=0}^{\infty} \sum_{s_z} \int_{-\infty}^{\infty} \frac{dk_{z}}{2\pi}\nn\\
 &~~~~~~~~~~~~~~~~~~~~~~\frac{\tau^{i}_R\omega_i(\Omega_{c_i}\tau^{i}_R)}{1+(\Omega_{c_i}\tau^{i}_R)^2}\left(\frac{{l|{q_i}|B}}{\omega_i^2}\right)
(-\frac{\partial f_i^{0}}{\partial \omega_i}).\label{equnew85}
 \end{align}
Hence, Landau quantization induces modifications to both leading- and higher-order thermoelectric coefficients.
\subsection{Higher-order thermoelectric coefficients in the presence of a time-varying magnetic field}
\label{formalism2}

The Boltzmann transport equation for a single particle species, within the RTA and under the influence of an external electromagnetic field, is mentioned in Eq.~\ref{equnew21}. Here, we consider a time-dependent electromagnetic field parameterized as~\cite{Satow:2014lia, Hongo:2013cqa}
\begin{align}
B = B_0 \exp\left(-\frac{\tau}{\tau_B}\right), \qquad
E = E_0 \exp\left(-\frac{\tau}{\tau_E}\right),
\label{Mag-Field}
\end{align}
where $B_0$ and $E_0$ denote the initial magnitudes of the magnetic and electric fields, respectively, while $\tau_B$ and $\tau_E$ are their decay parameters. The proper time is represented by $\tau$.
The exponential decay form of these fields follows naturally from Maxwell’s equations in a conducting medium, where finite electrical conductivity leads to temporal damping of the fields. Such forms typically emerge when Ohm’s law is incorporated into Maxwell’s equations. For simplicity, we focus on the temporal evolution of the fields in proper time $\tau$, neglecting spatial variations that are assumed to be smooth.

To solve Eq.~(\ref{equnew21}), we introduce an ansatz that characterizes the deviation from equilibrium as~\cite{Gavin:1985ph,Das:2021qii}
\begin{align}
\delta f_i = (\vec{k}_i \cdot \vec{\Omega}) \frac{\partial f^{0}_{i}}{\partial \omega_i},
\label{equnew22}
\end{align}
where the vector $\vec{\Omega}$ encodes the influence of the temperature gradient, electric field, and magnetic field. It can be expressed as
\begin{align}
\vec{\Omega} &= 
~\alpha_1 \vec{E} + \alpha_2 \dot{\vec{E}} + \alpha_3 \vec{\nabla}T + \alpha_4 (\vec{\nabla}T \times \vec{B})
+ \alpha_5 (\vec{\nabla}T \times \dot{\vec{B}}) \nonumber\\
&+ \alpha_6 (\vec{E} \times \vec{B})
+ \alpha_7 (\vec{E} \times \dot{\vec{B}})
+ \alpha_{8} (\dot{\vec{E}} \times \vec{B}),
\label{equnew23}
\end{align}
where $\alpha_j$ ($j=1,2,\ldots,8$) are unknown coefficients that determine the relative strength of each term in driving the system out of equilibrium.
All contributions consistent with the CP symmetries of the electric and heat currents are retained, including the time-varying electromagnetic field, temperature gradients, and their cross-terms up to the leading order.
In the case of a static electromagnetic field, the coefficients $\alpha_2$, $\alpha_5$, $\alpha_7$, and $\alpha_8$ vanish.
Furthermore, when the chiral chemical potential is zero, the $\vec{B}$ and $\dot{\vec{B}}$ terms do not contribute to the current~\cite{Satow:2014lia}.

Using the coefficients $\alpha_j$, the simplified form of $\delta f_i$ can be written as (see Appendix of Ref.~\cite{Singh:2024emy})
\begin{align}
\delta f_i &= \frac{-q_i \tau_R^i}{(1+\chi_i)(1+ \chi_i + \gamma_i^2)}
\Bigg[\Bigg((1+\chi_i)+\frac{\chi_i(1+ \chi_i - \gamma_i^2)}{(1+ \gamma_i^2)}\Bigg)\nonumber\\
&(\vec{v}_i \cdot \vec{E}) 
+ \Bigg(\gamma_i(1+\chi_i)
+\frac{\chi_i\gamma_i(2+ \chi_i)}{(1+ \gamma_i^2)}\Bigg)
\Big(\vec{v}_i \cdot (\vec{E}\times \hat{b})\Big)\Bigg] \nonumber\\
&+ \frac{(\omega_i - \beta_i h) \tau_R^i}{T(1+\chi_i)(1+ \chi_i + \gamma_i^2)}
\Bigg[(1+ \chi_i)\Big(\vec{v}_i\cdot\vec{\nabla}T\Big)\nonumber\\
&+\gamma_i(1+\chi_i)
\Big(\vec{v}_i\cdot(\vec{\nabla}T \times \hat{b})\Big)
\Bigg]\frac{\partial f^0_i}{\partial \omega_i}.
\label{equnew24}
\end{align}
For simplicity, we take $\tau_E = \tau_B$ and $\chi_i = \frac{\tau_R^i}{\tau_B} = \frac{\tau_R^i}{\tau_E}$. Here, $\gamma_i = \tau_R^i \Omega_{c_i}$, where $\Omega_{c_i} = \frac{q_iB}{\omega_i}$ is the cyclotron frequency of each hadron.

The induced electric current, obtained using $\delta f_i$, reads
\begin{align}
j^l &= \sum_{i} \frac{q_i g_i}{3} \int \frac{d^3\vec{k}_i}{(2\pi)^3} v_i^2
\frac{\tau_R^i}{(1+ \chi_i + \gamma_i^2)(1+\chi_i)} \nonumber\\
&\times \Bigg[-q_i\Bigg(
(1+\chi_i)
+\frac{\chi_i(1+ \chi_i - \gamma_i^2)}{(1+ \gamma_i^2)}\Bigg)\delta^{lk}E^k \nonumber\\
&+ \Bigg(\gamma_i(1+\chi_i)+
\frac{\chi_i\gamma_i(2+ \chi_i)}{(1+ \gamma_i^2)} \Bigg)
\epsilon^{ljk}h^jE^k 
+ \frac{(\omega_i - \beta_i h)}{T}\nonumber\\
&\Bigg((1+\chi_i)\delta^{lk}
\frac{\partial T}{\partial x^k}
+\gamma_i(1+\chi_i)\epsilon^{ljk}h^j
\frac{\partial T}{\partial x^k}\Bigg)\Bigg]
\frac{\partial f^0_i}{\partial \omega_i}.
\label{equnew36}
\end{align}
To simplify further, we choose the magnetic field along the $z$-axis, while the electric field and temperature gradient lie in the transverse ($x$–$y$) plane. Under these conditions, the components of the electric current in the plane are expressed as
\begin{widetext}
\begin{align}
j_x = & \sum_{i} \frac{q_i g_i}{3} \int \frac{d^3\vec{k}_i}{(2\pi)^3} v_i^2 
\frac{\tau_R^i}{(1+ \chi_i + \gamma_i)(1+\chi_i)}
\times \Bigg[-q_i\Bigg(
(1+\chi_i)
+\frac{\chi_i(1+ \chi_i - \gamma_i^2)}{(1+ \gamma_i^2)}\Bigg)E^x
+ \Bigg(\gamma_i(1+\chi_i)+
\frac{\chi_i\gamma_i(2+ \chi_i)}{(1+ \gamma_i^2)} \Bigg)
E^y\nonumber\\
&+ \frac{(\omega_i - \beta_i h)}{T}\Bigg((1+\chi_i)
\frac{\partial T}{\partial x}
+\gamma_i(1+\chi_i)
\frac{\partial T}{\partial y}\Bigg)\Bigg]\frac{\partial f^0_i}{\partial \omega_i},
\label{equnew91}
\end{align}
and
\begin{align}
j_y = & \sum_{i} \frac{q_i g_i}{3} \int \frac{d^3\vec{k}_i}{(2\pi)^3} v_i^2 
\frac{\tau_R^i}{(1+ \chi_i + \gamma_i^2)(1+\chi_i)} 
\times \Bigg[-q_i\Bigg(
(1+\chi_i)
+\frac{\chi_i(1+ \chi_i - \gamma_i^2)}{(1+ \gamma_i^2)}\Bigg)E^y
+ \Bigg(\gamma_i(1+\chi_i)-
\frac{\chi_i\gamma_i(2+ \chi_i)}{(1+ \gamma_i^2)} \Bigg)
E^x \nonumber\\
&+ \frac{(\omega_i - \beta_i h)}{T}\Bigg((1+\chi_i)
\frac{\partial T}{\partial y}
-\gamma_i(1+\chi_i)
\frac{\partial T}{\partial x}\Bigg)\Bigg]\frac{\partial f^0_i}{\partial \omega_i}.
\label{equnew92}
\end{align}

Equations~\eqref{equnew91} and~\eqref{equnew92} can be compactly rewritten by defining the following integrals:
\begin{align}
H_{1_i} &= \frac{g_i }{3} \int \frac{d^3\vec{k}_i}{(2\pi)^3}
\frac{\vec{k}_i^2}{\omega_i^2} ~(-\frac{\partial f_i^{0}}{\partial \omega_i}) \tau_R^i
\times \frac{(1+ \gamma_i^2)+\chi_i(2+ \chi_i)}{(1+\chi_i)(1+\gamma_i^2)(1+ \chi_i + \gamma_i^2)}, \label{equnew93}\\
H_{2_i} &= \frac{g_i}{3} \int \frac{d^3\vec{k}_i}{(2\pi)^3}
\frac{\vec{k}_i^2}{\omega_i^2} ~(-\frac{\partial f_i^{0}}{\partial \omega_i}) \tau_R^i
\times \gamma_i \frac{(1+\chi_i)(1+ \gamma_i^2)+\chi_i(2+ \chi_i)}{(1+\chi_i)(1+\gamma_i^2)(1+ \chi_i + \gamma_i^2)}, \label{equnew94}\\
H_{3_i} &= \frac{g_i}{3} \int \frac{d^3\vec{k}_i}{(2\pi)^3}
\frac{\vec{k}_i^2}{\omega_i^2}(\omega_i - \beta_i h)~(-\frac{\partial f_i^{0}}{\partial \omega_i}) \tau_R^i
\times \frac{1}{(1+\chi_i + \gamma_i^2)}, \label{equnew95}\\
H_{4_i} &= \frac{g_i}{3} \int \frac{d^3\vec{k}_i}{(2\pi)^3}
\frac{\vec{k}_i^2}{\omega_i^2}(\omega_i - \beta_i h)~(-\frac{\partial f_i^{0}}{\partial \omega_i}) \tau_R^i
\times \frac{\gamma_i}{(1+\chi_i + \gamma_i^2)}.
\label{equnew96}
\end{align} 
\end{widetext}
It is to be noted that these expressions are derived for the scenario of a time-varying magnetic field. For the case of a constant magnetic field ($\tau_B \rightarrow \infty$ or $\chi_i \rightarrow 0$ ), the expressions of components of electric currents in Eqs.~\eqref{equnew91}-~\eqref{equnew92} reduces to the Eqs.~\eqref{equnew38}-~\eqref{equnew39}.

\section{Results and Discussions}\label{sec-result}
 In this section, we discuss the results for higher-order thermoelectric coefficients of a hot and dense hadron gas in the presence of an external magnetic field. We also present a short description of the Thomson coefficient to show a comparative analysis between the magneto-Thomson coefficient and the Thomson coefficient. While the latter was estimated from the analysis of the Seebeck coefficient in our previous study~\cite{Singh:2025rwc}, the results shown here are obtained from more complete and rigorous calculations. This allows for a qualitative understanding of how the presence of a magnetic field modifies the higher-order transport properties, particularly in relation to the temperature dependence and baryon chemical potential. We have considered four different versions of hadron resonance gas models, including a discrete particle spectrum having all the hadrons and their resonances up to the mass cutoff $\Lambda$ = 2.6 GeV~\cite{ParticleDataGroup:2008zun}. To estimate the relaxation time of the hadrons, we follow the hard-scattering approximation. A uniform radius, $r_m$ = 0.2 fm for mesons and  $r_{b,\bar{b}}$ = 0.62 fm for baryons and antibaryons throughout the calculations ~\cite{Pradhan:2022gbm}. For the VDWHRG model, we take van der Waals parameters as $a$ = 0.926 GeV fm$^3$ and $b$ = (16/3)$\pi r^{3}$~\cite{Pradhan:2022gbm, Samanta:2017yhh}.

\label{results}
\begin{figure*}
	\centering
	\includegraphics[scale=0.27]{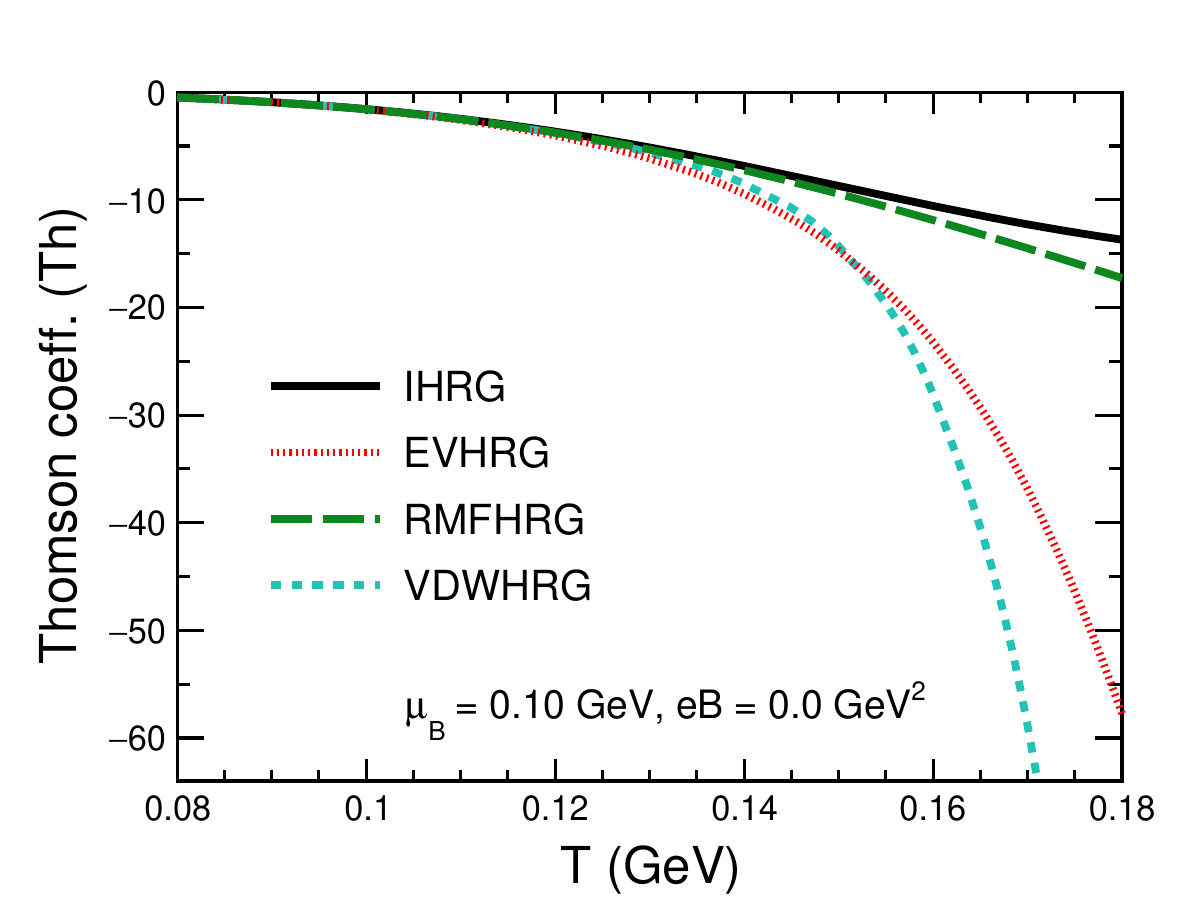}
 \includegraphics[scale=0.27]{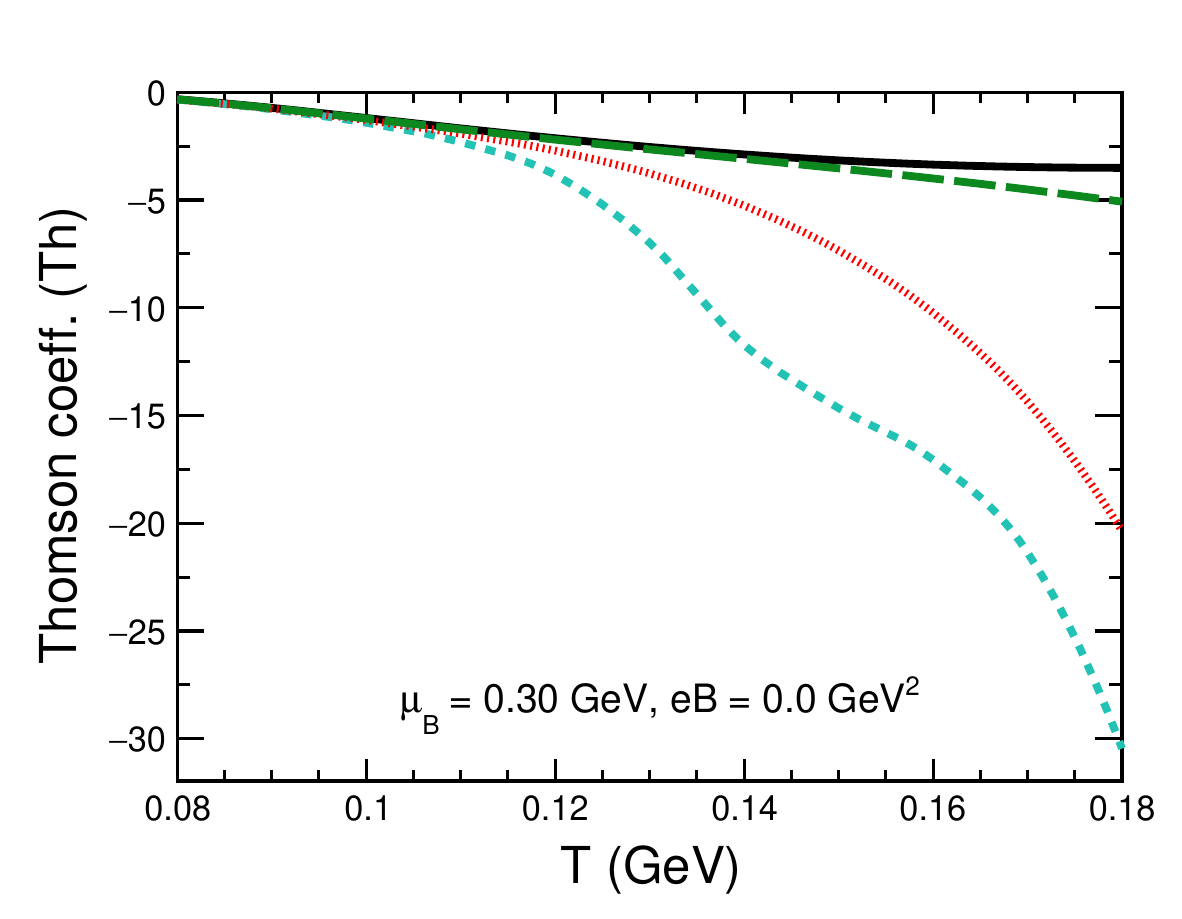}
  \includegraphics[scale=0.27]{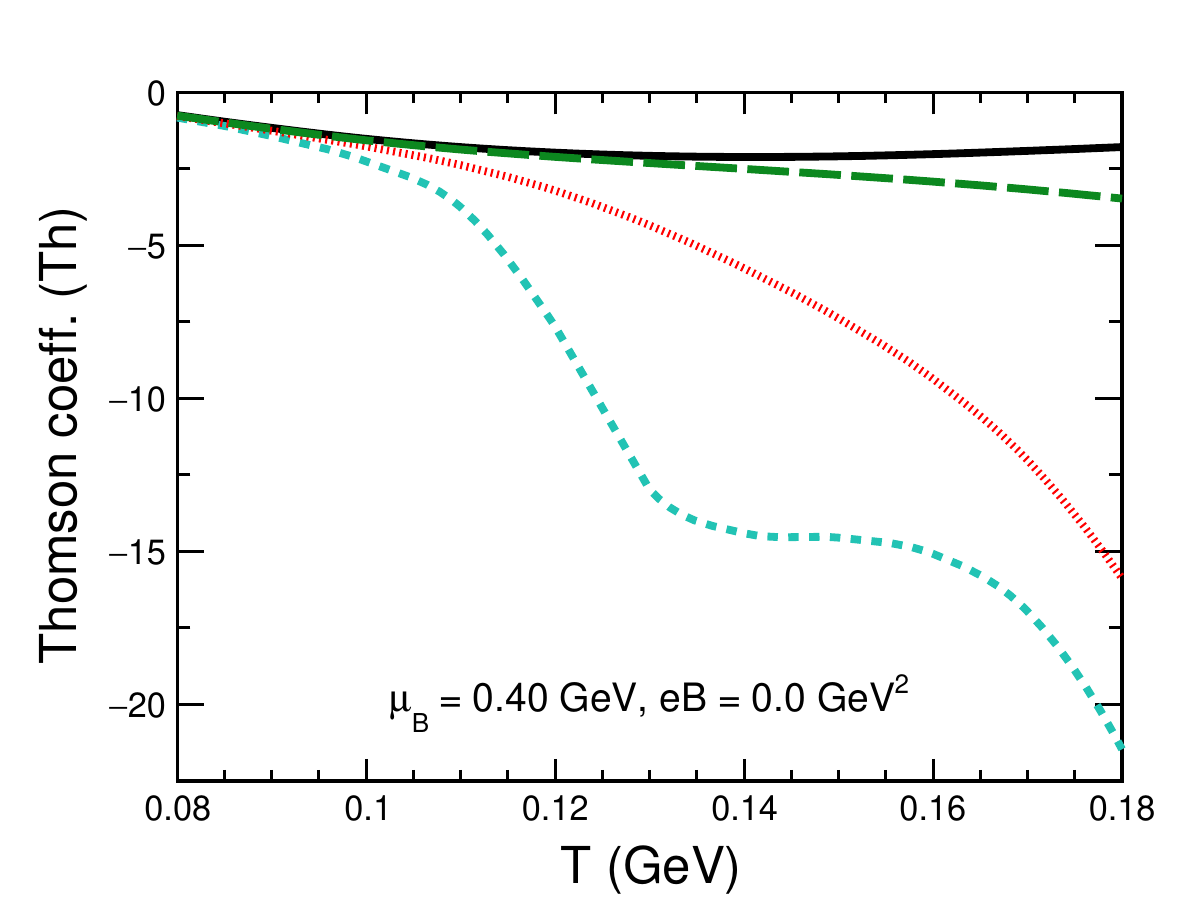}
	\caption{Thomson coefficient ($Th$) obtained in different hadronic models as a function of temperature at $\mu_{B}$ = 0.10 GeV (left panel), 0.30 GeV (middle panel), and 0.40 GeV (right panel).}
	\label{Fig-Thomson1}
\end{figure*}

\begin{figure*}
	\centering
	\includegraphics[scale=0.27]{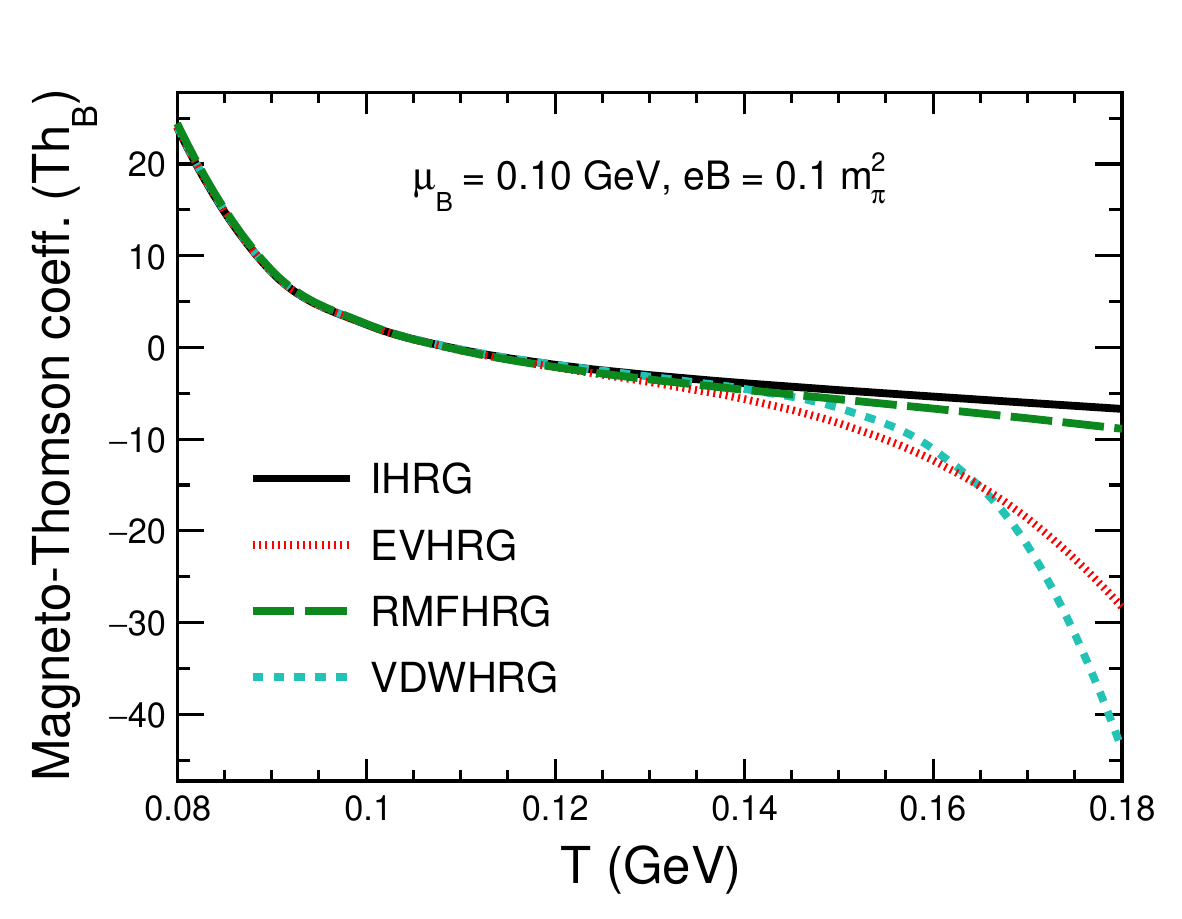}
    \includegraphics[scale=0.27]{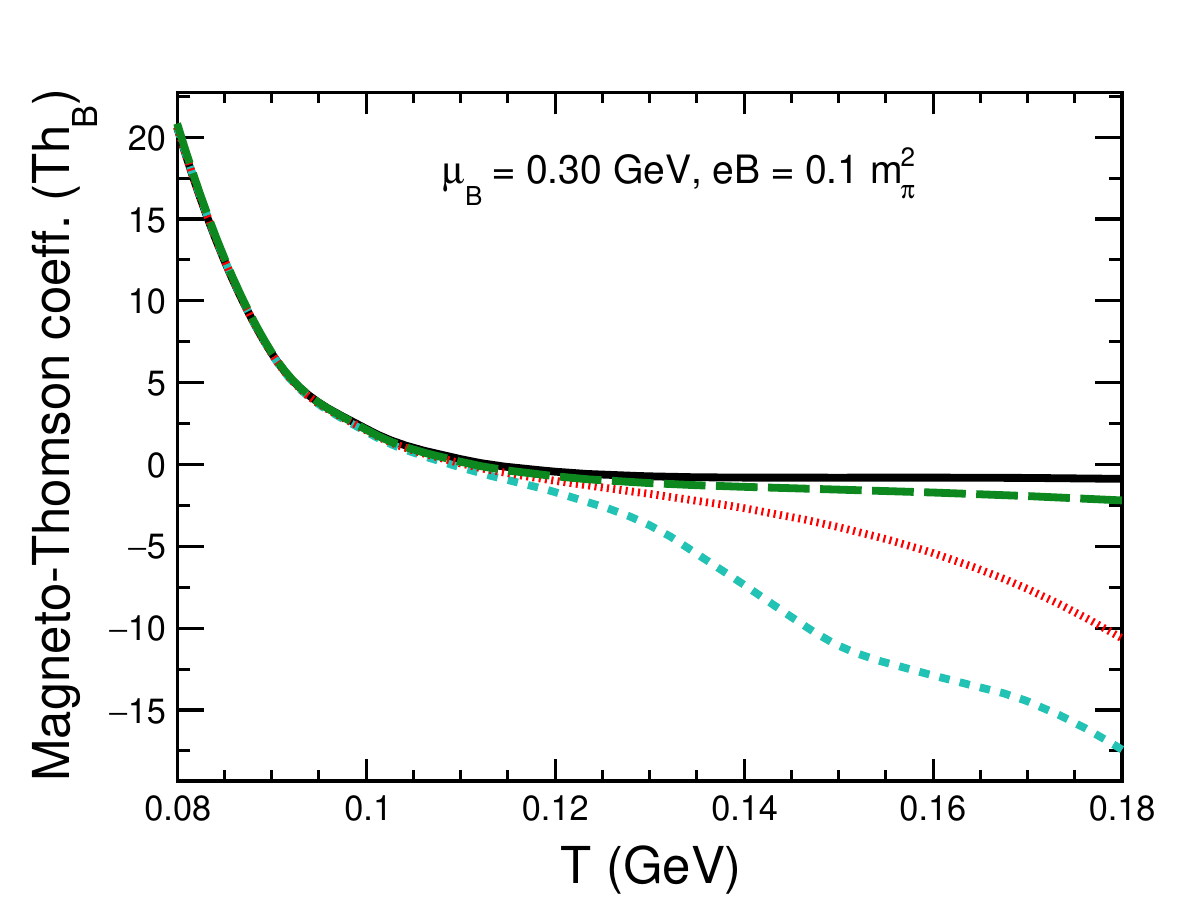}
    \includegraphics[scale=0.27]{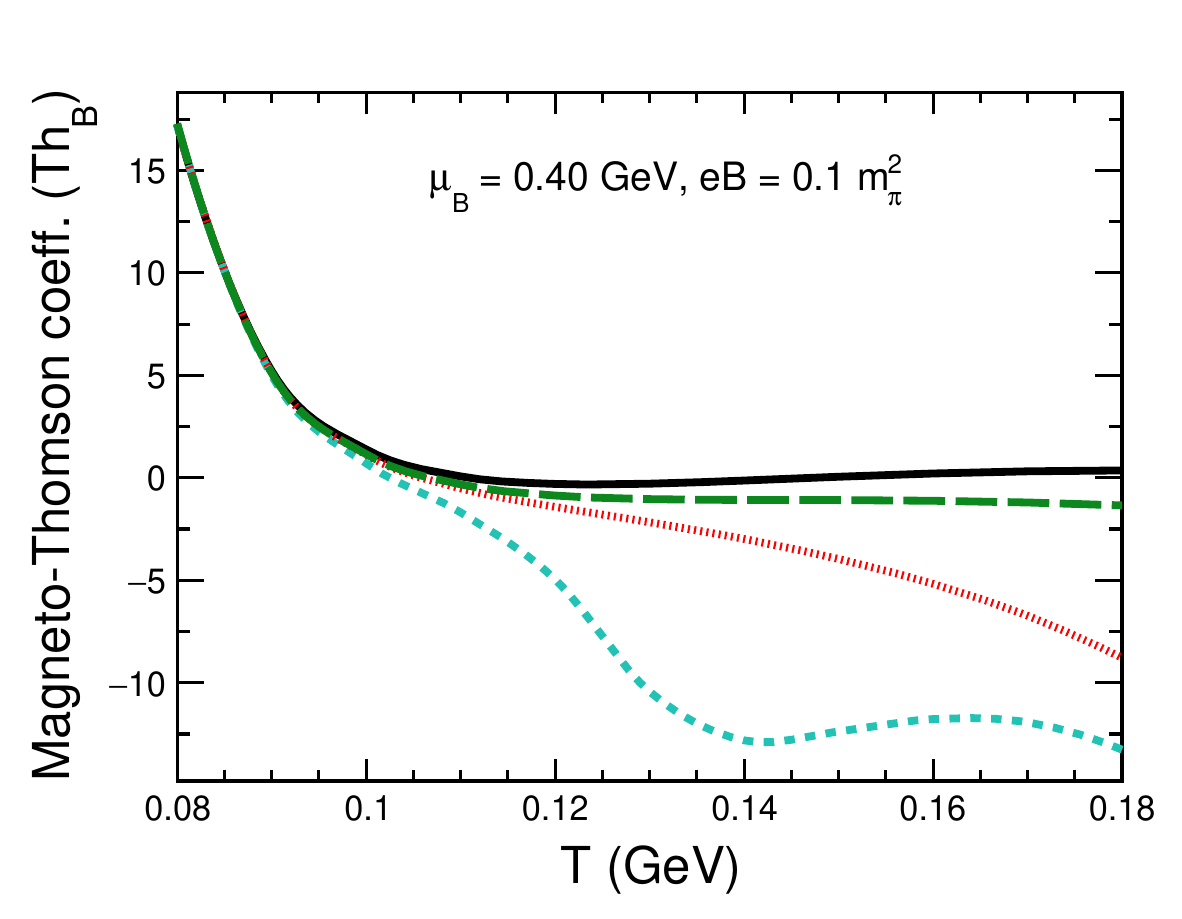}

	\includegraphics[scale=0.27]{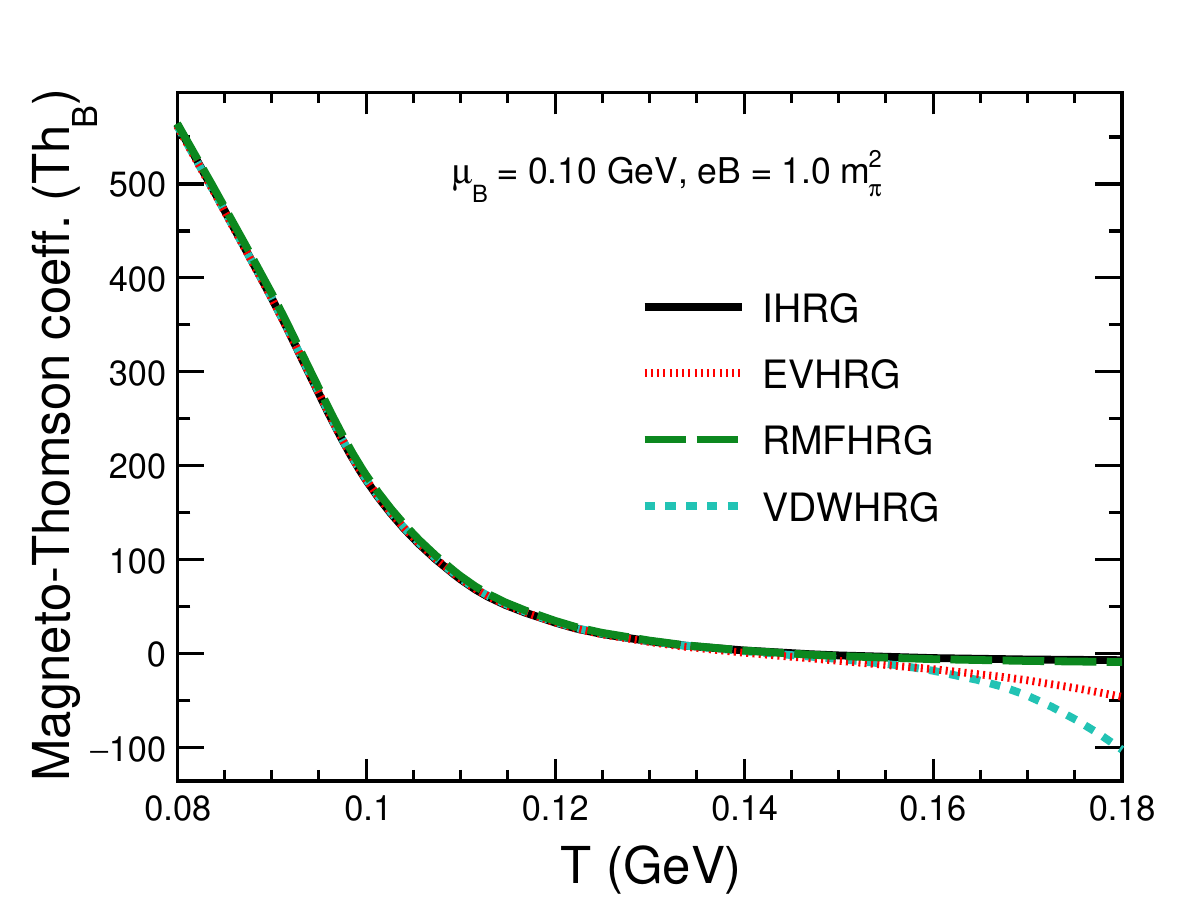}
    \includegraphics[scale=0.27]{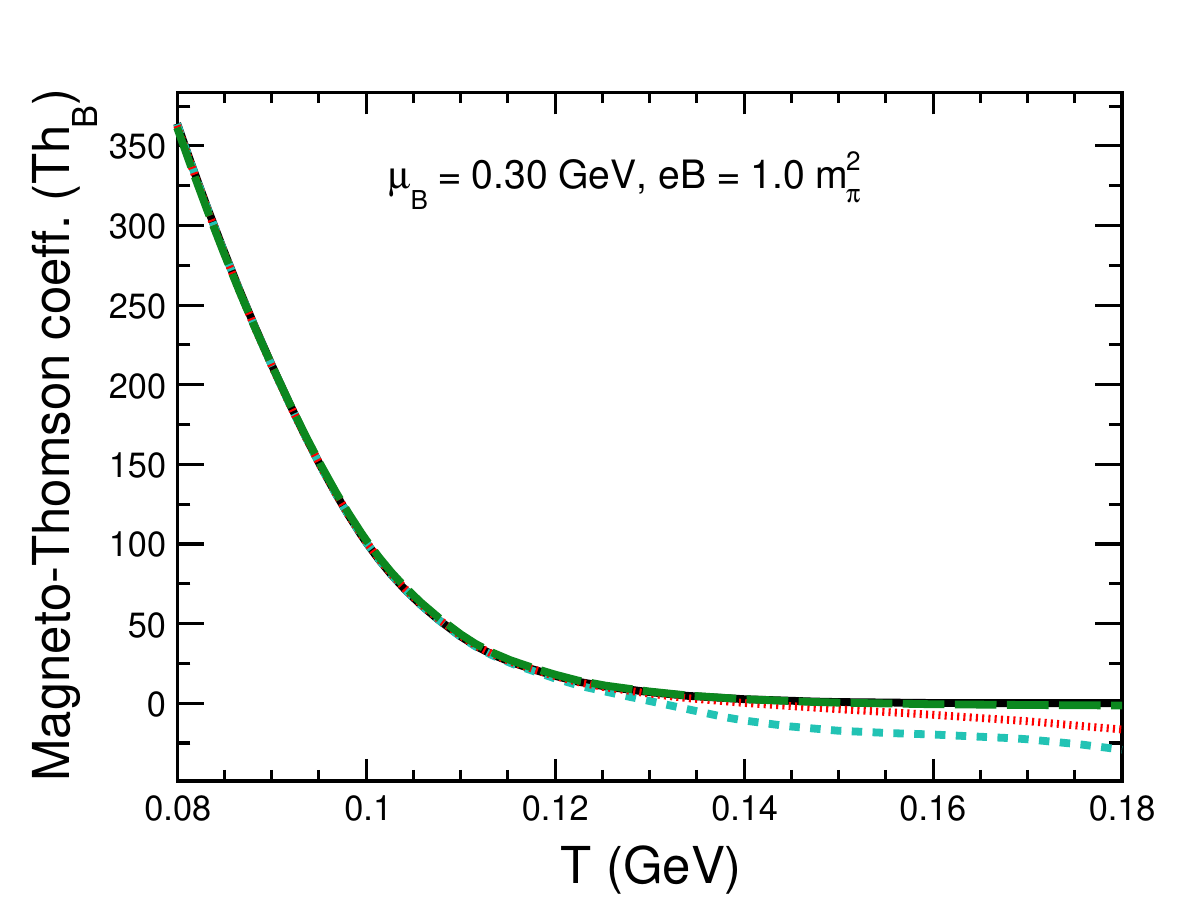}
    \includegraphics[scale=0.27]{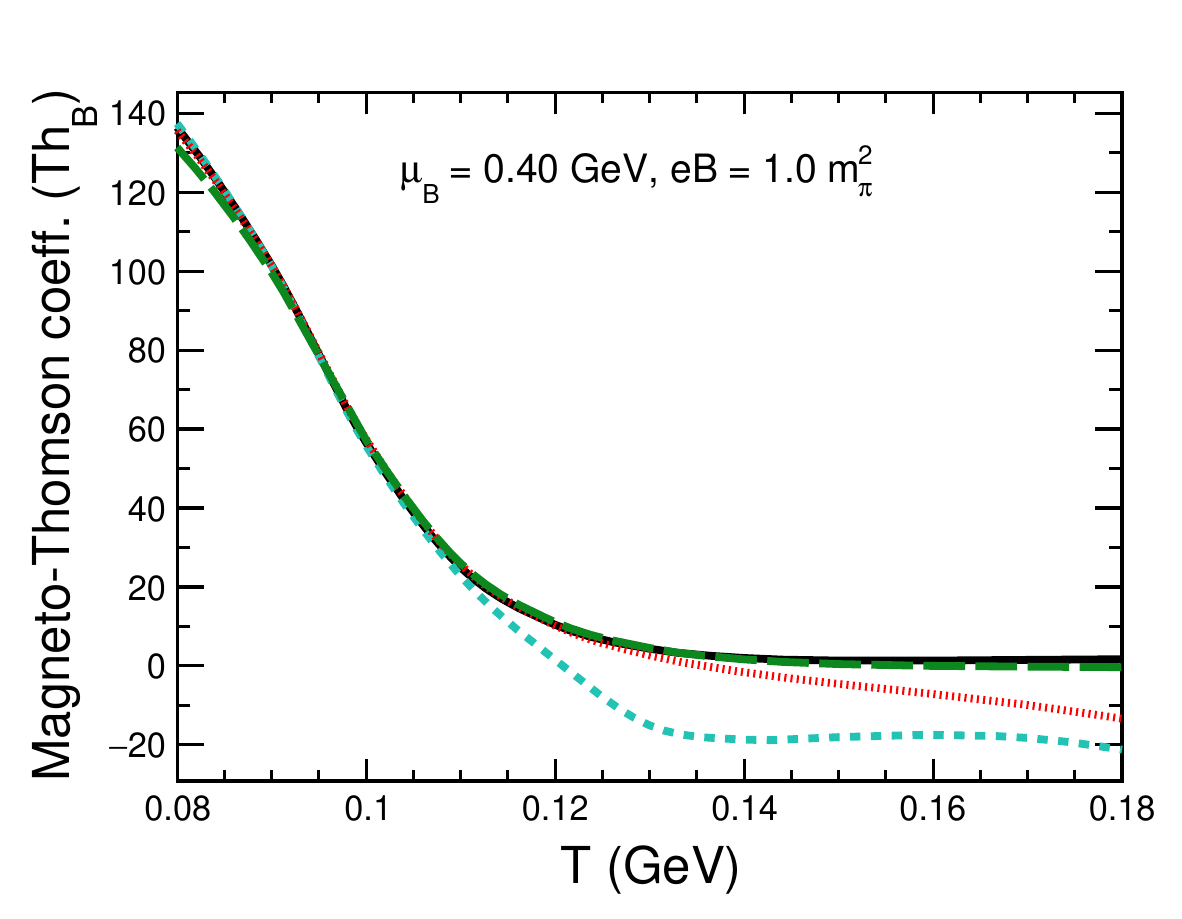}
    
	\caption{Magneto-Thomson coefficient ($Th_{B}$) as a function of temperature at magnetic field $eB$ = 0.1 $m_{\pi}^2$ (upper panel) and $eB$ = 1.0 $m_{\pi}^2$ (lower panel) for baryon chemical potential at $\mu_{B}$ = 0.10 GeV (left), 0.30 GeV (middle), and 0.40 GeV (right).}
	\label{Fig-magseebeck1}
\end{figure*}

Fig.~\ref{Fig-Thomson1} represents the Thomson coefficient $Th$ as a function of temperature $T$ for three different baryon chemical potentials, i.e., at $\mu_{B}$ = 0.10, 0.30, and 0.40 GeV. The black solid line represents the Thomson coefficient obtained in the IHRG model, whereas the red dotted line represents the EVHRG model. The green and cyan dashed lines represent the same for the RMFHRG and VDWHRG models, respectively. It is observed that for the chosen range of temperature (0.08 - 0.18 GeV), $Th$ is negative for a hot and dense hadron gas. The $Th$ is universally decreasing as the temperature increases. However, the degree of decrease is different for different models. To define
the net heat flow for a relativistic system, we need a conserved charge. For the study of the HRG system, the heat current corresponds to the net baryon current. It is observed that the baryons have a dominant contribution over the mesons to the Thomson coefficient. The mesons contribute to thermoelectricity only through the enthalpy of the system. In hadron gas, the entropy production for lighter mesons like pions, kaons, etc., is higher. Therefore, the enthalpy per baryon ($h$) exceeds the single-particle energy ($\omega_{i}$), which results in the negative values of $Th$~\cite{Singh:2025rwc}. The positive (negative) values of the Thomson coefficient lead to the net absorption (release) of heat in the system. Therefore,  a non-zero Thomson coefficient can significantly affect the cooling dynamics of the medium.  The Thomson coefficient can be calculated as the product of temperature and the rate of change of the Seebeck coefficient with respect to temperature~\cite{Singh:2025rwc}; therefore, the higher the rate of change of the  Seebeck coefficient with temperature, the higher the value of $Th$ in magnitude as a function of $T$. The coefficient $Th$ shows overlap at lower values of temperature, nearly up to $\approx 0.14$ GeV for the case of $\mu_B$ = 0.10 GeV. As the value of $\mu_B$ increases, the models start deviating from each other even at low temperatures. Additionally, a noticeable flatness appears in the VDWHRG model in the temperature range around $T\approx$ 0.13-0.16 GeV for $\mu_B$ = 0.40 GeV, indicating a plateau-like behavior of the Thomson coefficient in this region. On the other hand, the constant decreasing rate of the Seebeck coefficient around the above-mentioned temperature range~\cite{Singh:2025rwc} leads to the constant slope. Hence, the $Th$ shows a non-monotonic behavior at high values of $\mu_B$. 
The qualitative differences observed among the various HRG model results can be attributed to the manner in which different interaction mechanisms modify the fundamental thermodynamic inputs that enter transport phenomena. In this context, it is instructive to consider the behavior of the electrical conductivity scaled by temperature, $\sigma_{\mathrm{el}}/T$, since electrical conductivity provides the leading contribution to thermoelectric transport coefficients and therefore serves as a useful baseline for understanding model-dependent trends.
At higher temperatures, the number density of the medium increases rapidly, and the repulsive component present in interacting HRG models becomes increasingly significant, as discussed in the literature~\cite{Pradhan:2022gbm}. This repulsive interaction effectively reduces the number of particles per unit volume, which in turn enhances the relaxation time~\eqref{equnew70} relative to the IHRG scenario. As a consequence, a finite enhancement in the magnitude of $\sigma_{\mathrm{el}}/T$ and similarly in other transport coefficients is observed once hadronic interactions are incorporated. With increasing baryon chemical potential $\mu_B$, the baryon number density further increases, causing the deviations from the IHRG behavior to shift toward lower temperatures.
Since the Thomson coefficients are constructed from the same underlying thermodynamic quantities, such as number density and enthalpy, as well as the same relaxation times, their model-dependent behavior naturally follows the same physical mechanism. This explains why differences among HRG models become more pronounced in regimes of higher temperature and baryon density.

\begin{figure*}
	\centering
	\includegraphics[scale=0.27]{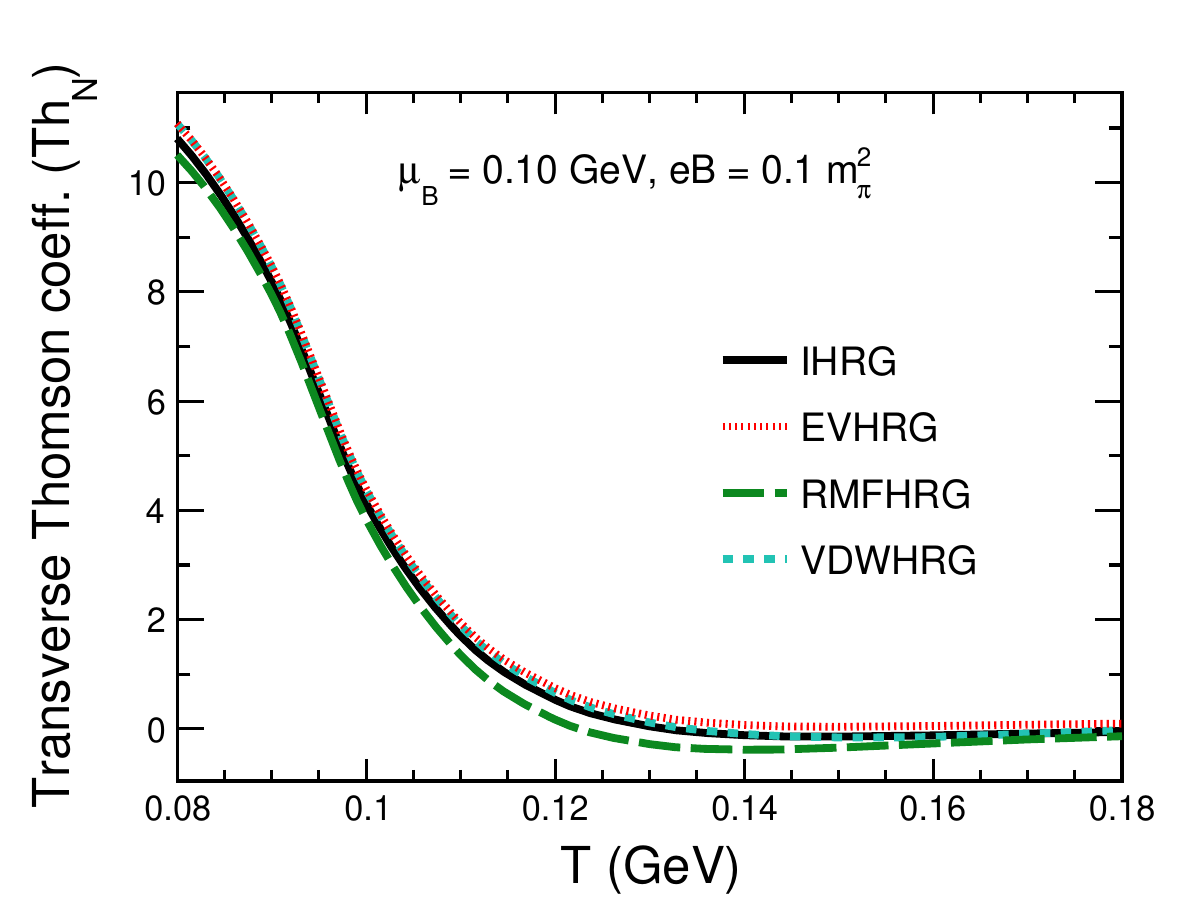}
    \includegraphics[scale=0.27]{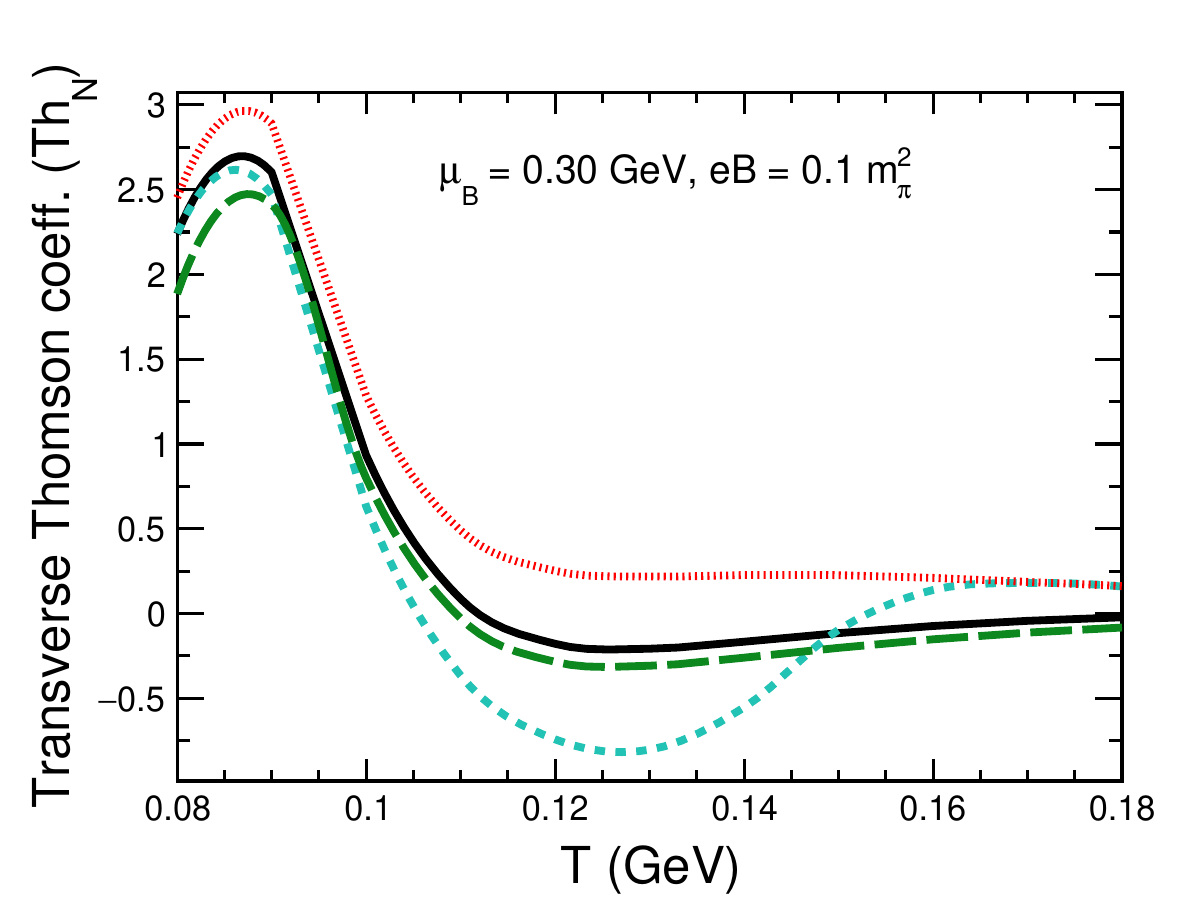}
    \includegraphics[scale=0.27]{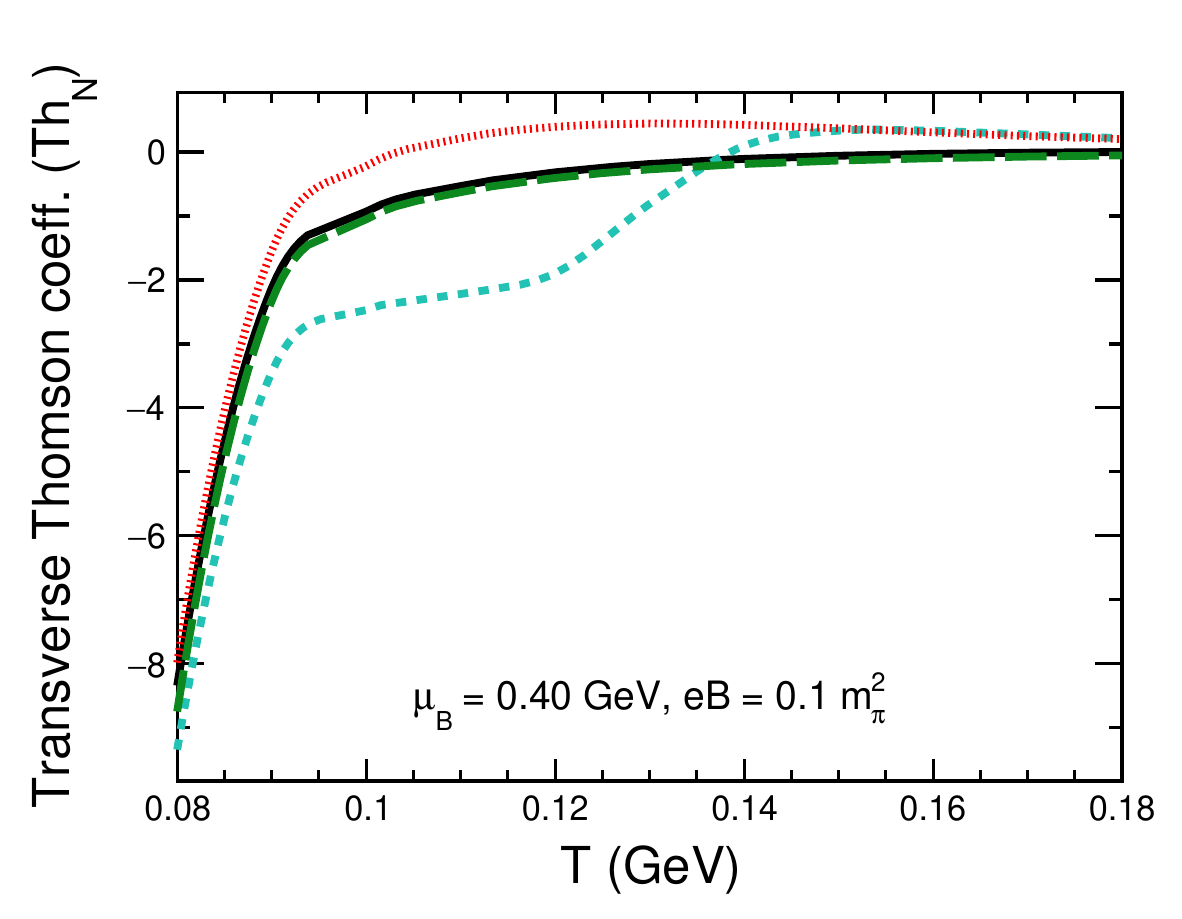}

	\includegraphics[scale=0.27]{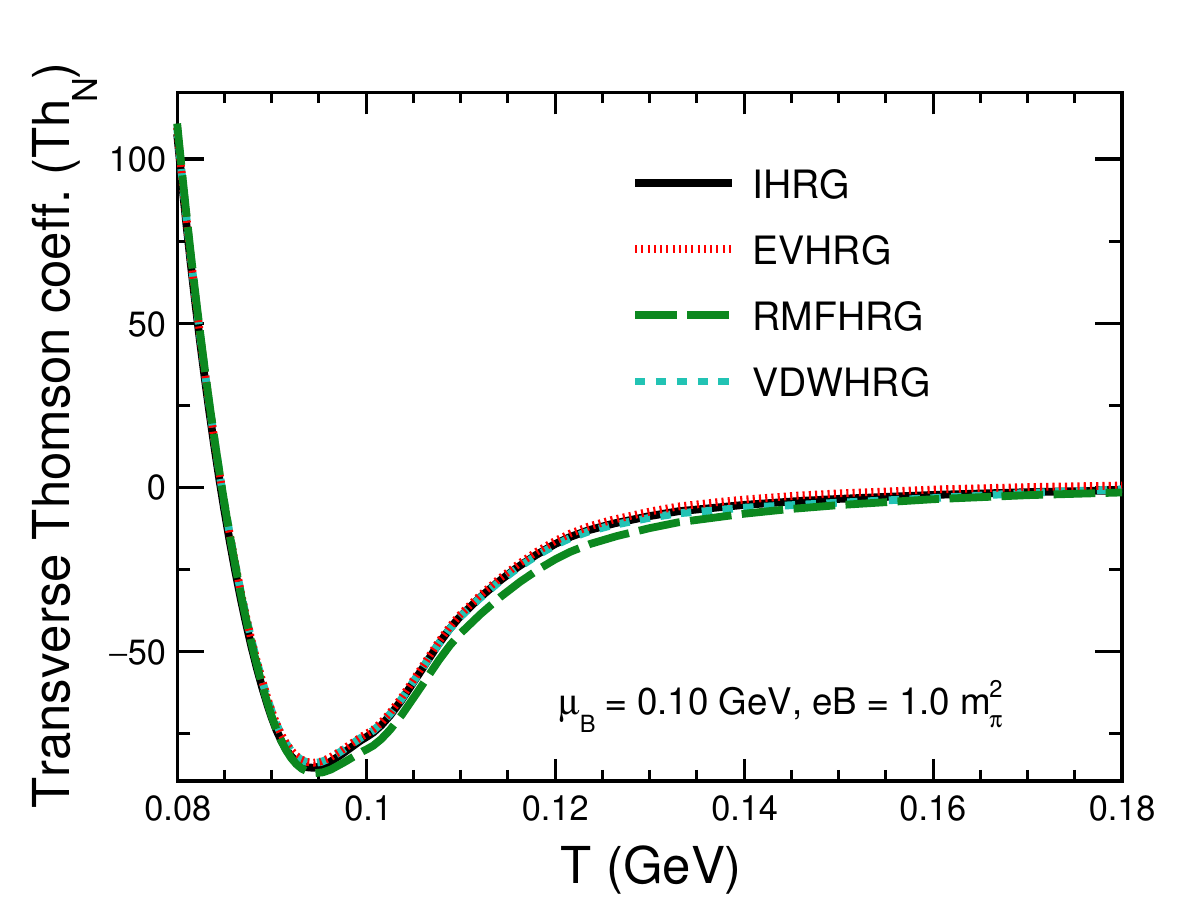}
    \includegraphics[scale=0.27]{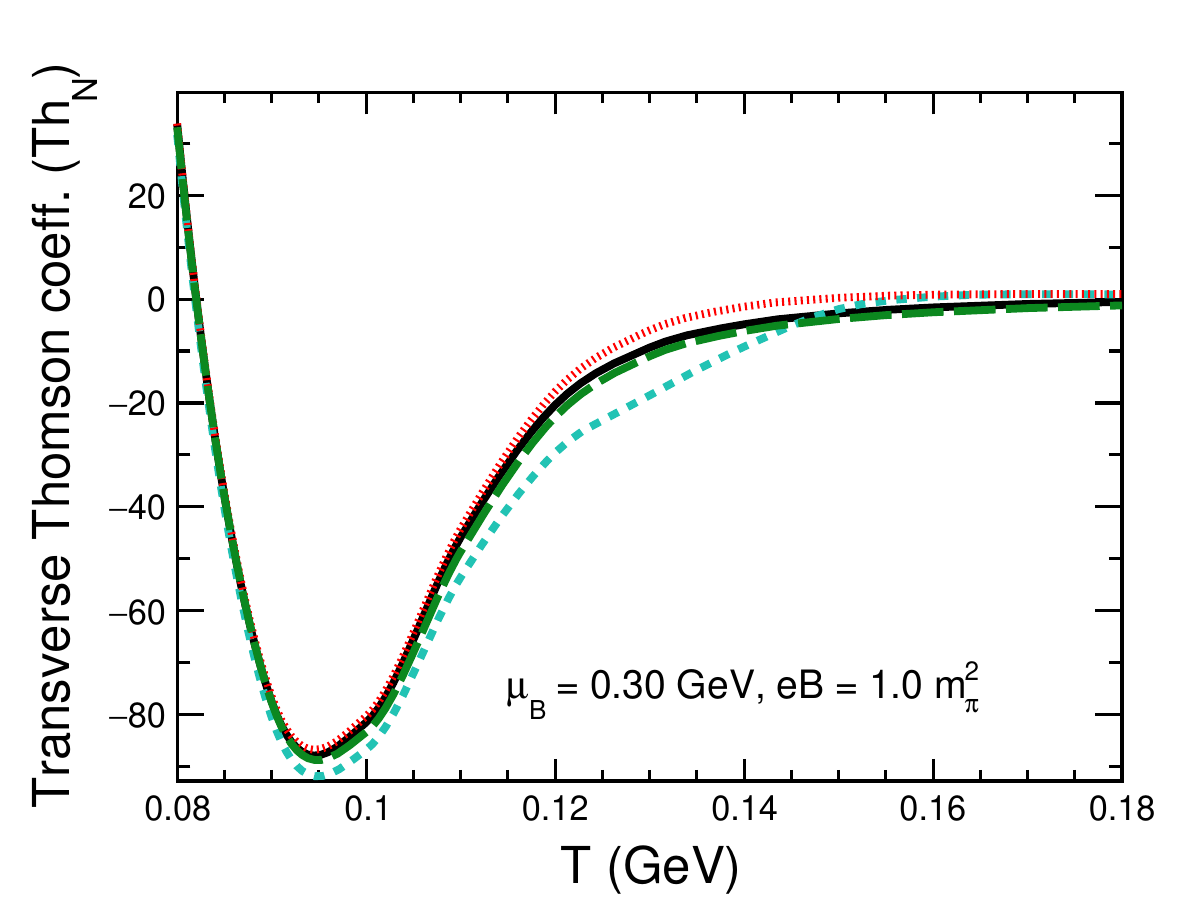}
    \includegraphics[scale=0.27]{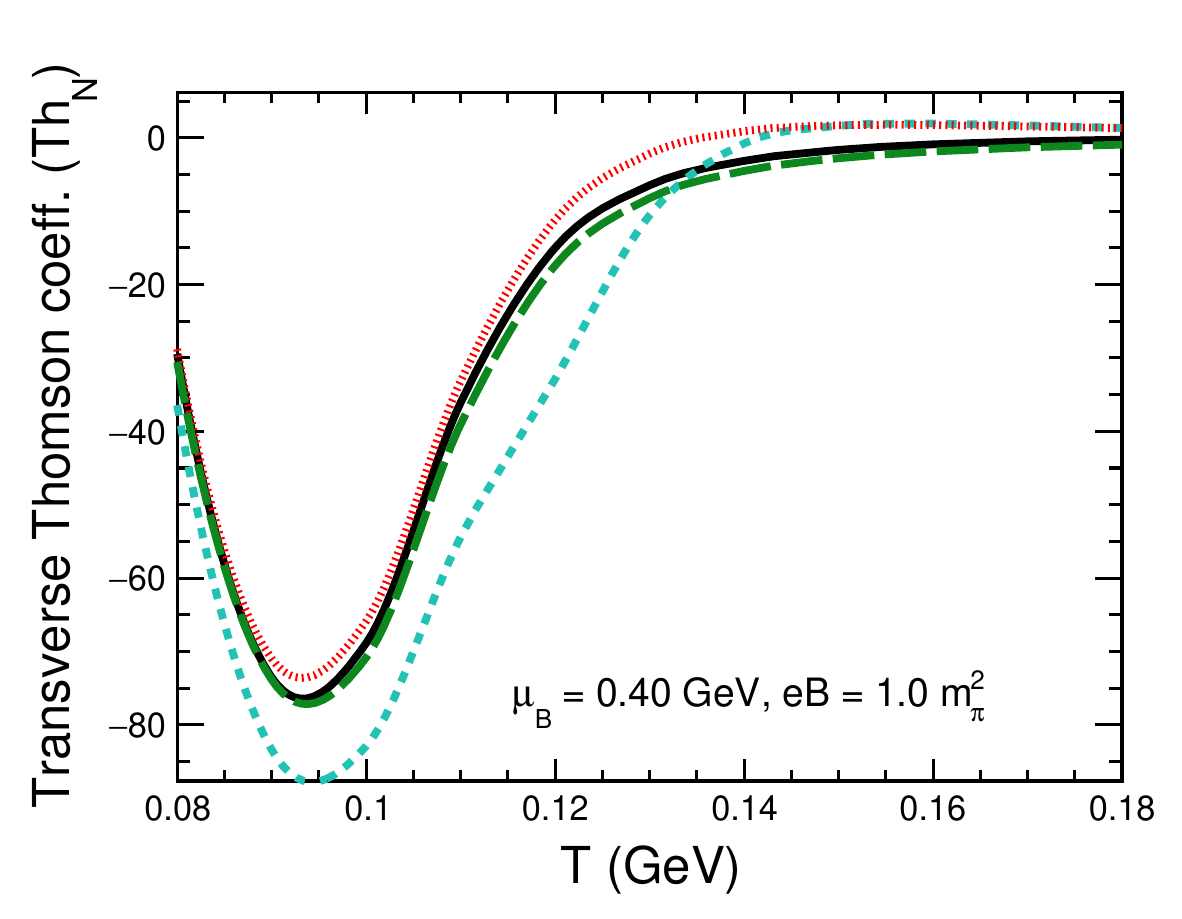}
	\caption{Transverse Thomson coefficient ($Th_N$) as a function of temperature at magnetic field $eB$ = 0.1 $m_{\pi}^2$ (upper panel) and $eB$ = 1.0 $m_{\pi}^2$ (lower panel) for baryon chemical potential at $\mu_{B}$ = 0.10 GeV (left), 0.30 GeV (middle), and 0.40 GeV (right).}
	\label{Fig-nernst1}
\end{figure*}

 The upper panel of Fig.~\ref{Fig-magseebeck1} represents $Th_{B}$ as a function of $T$ for three different values of $\mu_{B}$ at  0.10, 0.30, and 0.40 GeV at magnetic field 0.1 $m_{\pi}^2$, whereas the lower panel represents those for the magnetic field 1.0 $m_{\pi}^2$. For the case of $eB$ = 0.1 $m_\pi^2$, we observe that $Th_{B}$ decreases with the temperature up to 0.1 GeV, becomes zero at a particular $T$, and further, it starts to increase in the negative direction for higher values of temperature. Similar to what was observed in the case of a vanishing magnetic field, all the models agree up to a certain temperature and then start deviating from each other at higher temperatures. Furthermore, we also observe that the values of $Th_{B}$ increase in the negative direction with the increasing values of $\mu_{B}$ for higher $T$ but decrease in the positive direction for the lower values of $T$. Unlike the case without a magnetic field, the magneto-Thomson coefficient $Th_{B}$ exhibits an additional dependence on the Hall-like component of electrical conductivity and other integral terms that arise due to the presence of an external magnetic field. The Hall-like conductivity term, $\sigma_H/T$, increases with baryon chemical potential $\mu_B$ because of the growing net baryonic contributions. It is important to note that mesons do not contribute directly to the integrals $\mathcal{I}_{31}/T^2$ and $\mathcal{I}_{42}/T^2$, though their influence appears indirectly through the enthalpy per baryon $h$. The effect of the magnetic field is significant on $\mathcal{I}_{31}/T^2$ at low temperatures, but almost diminishes at higher temperatures due to a reduced relaxation time. On the other hand, $\mathcal{I}_{42}/T^2$ first increases with temperature and then starts to decrease. The higher the magnetic field, the lower the $\mathcal{I}_{42}/T^2$.  Here, the magnitude of $\sigma_{el}/T$ is significantly larger than that of $\sigma_H/T$, the integrals $\mathcal{I}_{31}/T^2$ and $\mathcal{I}_{42}/T^2$ are of comparable scale~\cite{Das:2019wjg}. Consequently, the magneto-Seebeck coefficient $S_B$ can be approximately represented by the ratio $(\mathcal{I}_{31}/T^2)/(\sigma_{el}/T)$, and its temperature dependence is reflected in the behavior of the magneto-Thomson coefficient $Th_B$, as shown in Fig.~\ref{Fig-magseebeck1}. The $T$ and $\mu_B$ dependence of $\sigma_{el}/T$ in a magnetic field causes $S_B$ to increase at low $T$ and decrease at high $T$. This results in a positive $Th_B$ at lower temperatures, which decreases with $T$ and becomes negative at higher temperatures. At low $T$, a large value of $\Omega_{c_i} \tau_{R_i}$ suppresses the mesonic contribution to $\sigma_{el}/T$, and this reduction is compensated by the enhanced baryonic contribution as $\mu_B$ increases. However, at higher temperatures, the baryonic increase is insufficient to compensate for the mesonic suppression, leading to an overall decrease in $\sigma_{el}/T$ with $\mu_B$. For stronger magnetic fields, as shown in the lower panel of Fig.~\ref{Fig-magseebeck1}, $Th_B$ shows large and positive values at low $T$, gradually approaching zero at high $T$. It is observed that for higher values of a magnetic field, results from all the models agree up to $T\approx0.12$ GeV at $\mu_B$ = 0.10 GeV. With increasing values of $\mu_B$, the disagreement between the models starts from even lower values of $T$. This observation is in line with what was observed for the magneto-Seebeck coefficient in Ref.~\cite{Singh:2025rwc}.
\begin{figure*}
	\centering
	\includegraphics[scale=0.37]{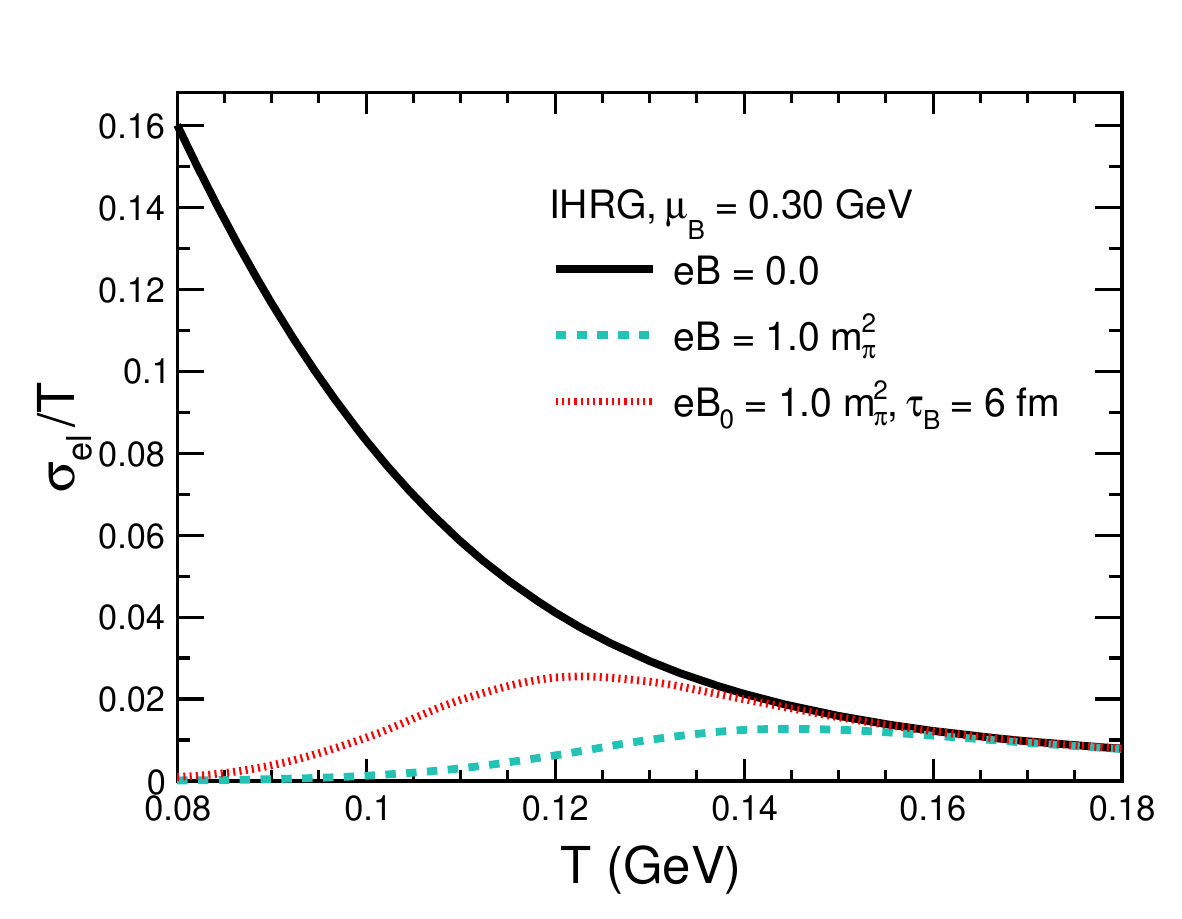}
 \includegraphics[scale=0.37]{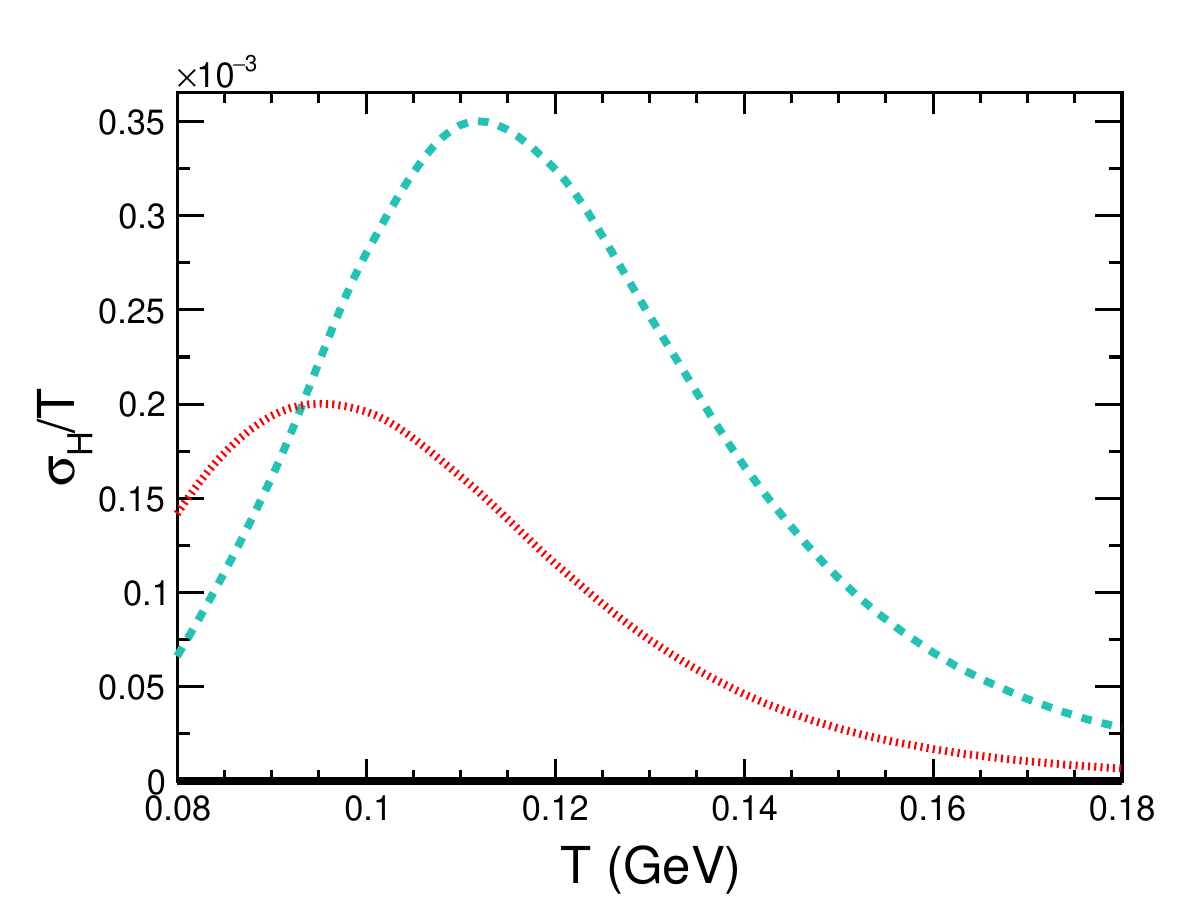}
  \includegraphics[scale=0.21]{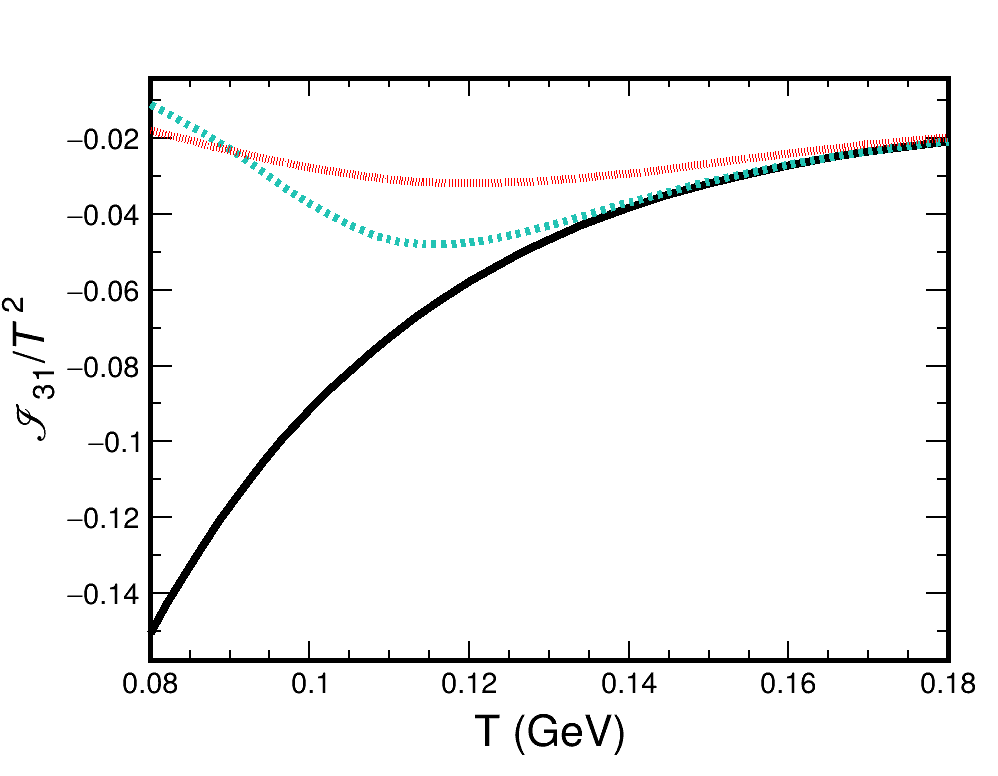}
  \includegraphics[scale=0.21]{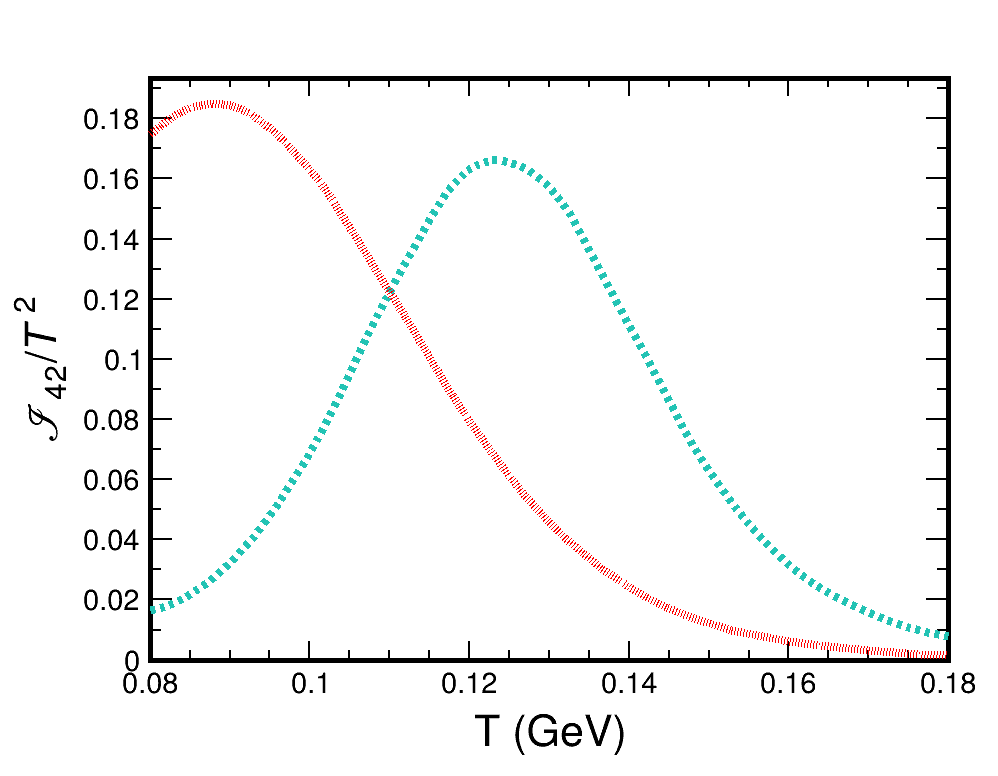}
	\caption{Upper panel: left figure shows scaled electrical conductivity ($\sigma_{el}/T$) and right figure shows scaled Hall-like component of electrical conductivity ($\sigma_{H}/T$), bottom panel: left figure shows scaled integral ($\mathcal{I}_{31}/T^2$) and right figure shows scaled  integral ($\mathcal{I}_{42}/T^2$) at $\mu_{B}$ = 0.30 GeV with three different scenario of magnetic field $eB$ = 0.0, $eB$ = 0.1 $m_{\pi}^2$, and $eB_0$ = 1.0 $m_{\pi}^2$ with decay parameter $\tau_B$ = 6 fm.}
	\label{Fig-conductivity}
\end{figure*}

\begin{figure*}
	\centering
	\includegraphics[scale=0.37]{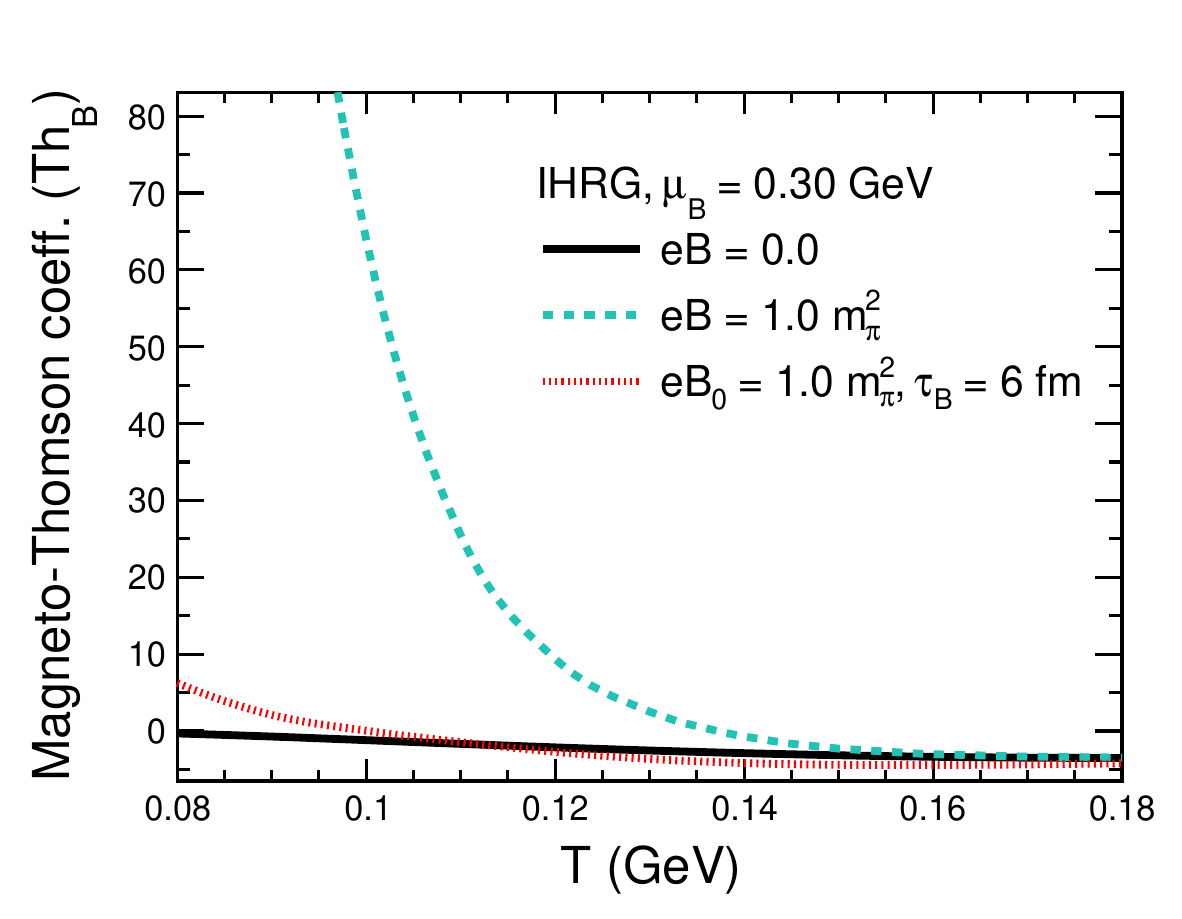}
    \includegraphics[scale=0.37]{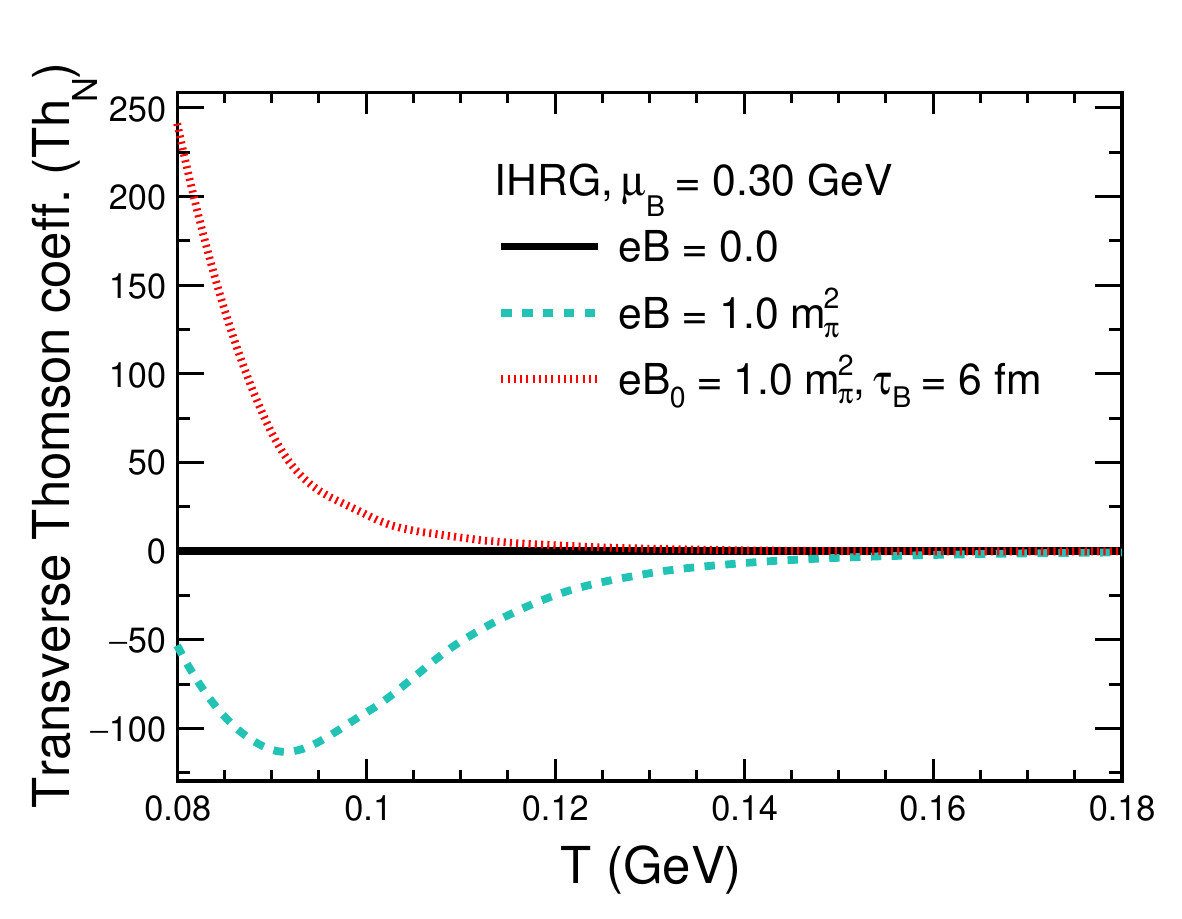}
    
	\caption{Left panel: Magneto-Thomson coefficient ($Th_{B}$), and right panel: transverse Thomson coefficient ($Th_{N}$)  as a function of temperature for baryon chemical potential at $\mu_{B}$ = 0.30 GeV with three different scenario of magnetic field $eB$ = 0.0, $eB$ = 0.1 $m_{\pi}^2$, and $eB_0$ = 1.0 $m_{\pi}^2$ with decay parameter $\tau_B$ = 6 fm.} 
	\label{Fig-ThomsonBt}
\end{figure*}
In the presence of an external magnetic field, we also observe a Hall-like component for the higher-order thermoelectric coefficients. The presence of a non-zero temperature gradient in a conducting medium directed perpendicular to the direction of a magnetic field gives rise to the Nernst coefficient. It measures the electric current perpendicular to both the magnetic field and the temperature gradient. In the absence of a magnetic field, this coefficient vanishes. The temperature dependence of the Nernst coefficient in the medium further gives rise to the transverse Thomson coefficient $Th_N$ as shown in Eq.~\eqref{Trans-Thomson}. Fig.~\ref{Fig-nernst1} represents $Th_N$ as a function of $T$ for three different values of $\mu_{B}$ at  0.10, 0.30, and 0.40 GeV. The upper and lower panels present the results for the case of the magnetic field with values 0.1 and 1.0 $m_{\pi}^2$, respectively. In the upper panel, we observe that $Th_N$ is positive for lower values of $T$ up to 0.10 GeV for $\mu_B$ = 0.10 and 0.30 GeV. The positive values of $Th_N$ signify that in Eq.~(\ref{Trans-Thomson}), the quantity 2$NB$ is dominating over the dynamical term, $T \frac{dNB}{dT}$ at lower temperatures. The impact of this dynamic part slowly starts increasing with $T$, and hence the negative behavior of $Th_N$ is also observed at higher $T$. For the higher values of $\mu_{B}$, it is observed that the dynamical part of Eq.~\eqref{Trans-Thomson} is dominating almost for the whole temperature range considered here. As it is observed in Ref.~\cite{Singh:2025rwc}, the coefficient $NB$ vanishes near high temperatures; here, $Th_N$ also starts disappearing at higher temperatures.
For $eB$ = 1.0~$m_{\pi}^2$, results from different models are observed to be almost similar, \i.e., no significant model dependence is observed. Even at higher $\mu_B$, only a slight deviation in the models can be seen. As one goes to a higher value of the magnetic field ($eB$ = 1.0~$m_\pi^2$), $Th_N$ is observed to be higher. Due to the higher order of magnitude of $\sigma_{el}/T$ as compare to $\sigma_{H}/T$, $NB$ can be approximated by ratio $(\mathcal{I}_{42}/T^2)/(\sigma_{el}/T)$. The sharp decrease of $\mathcal{I}_{42}/T^2$ as compare to $\sigma_{el}/T$ give rise to decreasing trend of $NB$ with $\mu_{B}$. Though $NB$ decreases with increasing $\mu_B$ but the dynamical rate of decrease of $NB$ increases. Hence, the coefficient $Th_N$ mostly has negative values throughout the whole temperature range at high values of $\mu_B$. Here, a clear effect of magnetic field values can be observed in the upper and bottom panels of Fig.~\ref{Fig-nernst1}. For the case of a higher magnetic field, the effect of the Lorentz force is also high. Thus, the higher Lorentz force increases the thermoelectric effects in the transverse direction, which results in higher values of the transverse Thomson coefficient. Once this magnetic field vanishes, the coefficient $Th_N$ drops to zero only.  

For completeness, we further extend this study to the case of a time-varying electromagnetic field. Here, as mentioned in Eq.~\ref{Mag-Field}, we have considered the exponential profile of the electromagnetic field. As the hadronic model considered in this study represents a static medium, we observe the effects of a time-varying magnetic field in a snapshot at proper time $\tau$ = 9 fm. The results presented in Figs.~\ref{Fig-conductivity} and~\ref{Fig-ThomsonBt} are focused on examining the influence of a time-varying magnetic field on the transport behavior of the hadronic medium. As the magnetic field decays and weakens over time, the effects of Landau quantization become negligible. Therefore, for a consistent comparison between the dynamical and static magnetic field pictures, Landau quantization has not been included in either case in these plots, allowing us to keep the focus on the dynamical effects of the temporal evolution of the field. As shown in Fig.~\ref{Fig-conductivity}, the upper-left panel displays the variation of the scaled electrical conductivity $\sigma_{el}/T$ with temperature, while the upper-right panel presents the Hall-like component $\sigma_{H}/T$. The lower panels correspond to the scaled heat current integrals $\mathcal{I}_{31}/T^{2}$ and $\mathcal{I}_{42}/T^{2}$, which constitute the underlying inputs for evaluating the leading-order and higher-order thermoelectric coefficients. Three distinct scenarios are compared: the case without a magnetic field ($eB=0.0$, black solid line), a constant magnetic field ($eB = 1.0\,m_{\pi}^{2}$, cyan dashed line), and a time-dependent magnetic field with decay parameter $\tau_{B} = 6~\mathrm{fm}$ (red dotted line).
At zero magnetic field, $\sigma_{el}/T$ exhibits a monotonically decreasing trend with increasing temperature, consistent with the expected reduction of medium interaction effects at higher temperatures~\cite{Das:2020beh}. The introduction of a magnetic field modifies the charge transport behavior, leading to an overall suppression in $\sigma_{el}/T$ at low temperatures due to the cyclotron motion of charged hadrons. The damping effect for the case of a decaying magnetic field is smaller due to the reduction in the magnitude of the magnetic field. In contrast, the Hall-like component $\sigma_{H}/T$ shows a non-monotonic structure, attaining a peak in the low-temperature regime where the Lorentz force contribution is dominant. The peak shifts towards lower temperatures and decreases in magnitude when the field becomes time dependent, indicating a dynamical competition between the magnetic relaxation time and the microscopic relaxation processes.

The thermal integrals $\mathcal{I}_{31}/T^{2}$ and $\mathcal{I}_{42}/T^{2}$ exhibit complementary trends. The integral $\mathcal{I}_{31}/T^{2}$ remains negative throughout the temperature range, gradually approaching zero as the temperature increases, while $\mathcal{I}_{42}/T^{2}$ shows a pronounced peak structure analogous to that of $\sigma_{H}/T$. These behaviors underline the sensitivity of heat flow to both the strength and temporal variation of the magnetic field. Overall, the comparison among the three cases shows that the presence of a time-varying magnetic field reduces the magnitude of both electrical and thermal transport in the hadronic medium. The decaying field weakens the response of charged particles to the electromagnetic force, resulting in smoother and less pronounced variations of the transport coefficients with temperature. This suggests that time-dependent magnetic effects can significantly impact the charge and heat transport properties of the hadronic matter produced in heavy-ion collisions, providing a more realistic description of the medium in evolving magnetic backgrounds.

The results presented in Fig.~\ref{Fig-ThomsonBt} show the temperature dependence of the magneto-Thomson coefficient $Th_B$ and the transverse Thomson coefficient $Th_N$ for different magnetic field configurations in the hadronic medium at $\mu_B = 0.30~\text{GeV}$. The integrals $H_{1_i}$-~$H_{4_i}$, shown in Eqs.~\ref{equnew93}-~\ref{equnew96}, encapsulate the effects of electric and heat current responses in the presence of both constant and time-dependent magnetic fields, and serve as the essential building blocks for evaluating magneto-Seebeck coefficient $S_B$ and normalized Nernst coefficient $NB$.
The left panel shows that $Th_B$ decreases rapidly in magnitude with temperature for all cases, indicating a diminishing magneto-thermoelectric response in the high-temperature regime. At zero magnetic field, $Th_B$ remains relatively constant and small in magnitude, while the introduction of a finite magnetic field ($eB = 1.0\,m_\pi^2$) enhances the response significantly at lower temperatures. When the magnetic field is allowed to decay in time with a decay parameter $\tau_B = 6$~fm, the enhancement is notably suppressed, leading to a less pronounced temperature dependence. This demonstrates that the temporal damping of the magnetic field reduces the magnetically induced thermoelectric effects in the hadronic medium.
The right panel shows $Th_N$, which reflects the Hall-like component of higher-order thermoelectric transport. Its temperature dependence arises from the interplay between the temperature derivative of the normalized Nernst coefficient and its magnitude. In the absence of a magnetic field, $Th_N$ remains zero across the entire temperature range due to the absence of $NB$. When a static magnetic field is introduced, $Th_N$ attains a negative value in the lower temperature region, indicating the dynamical part $T\frac{dNB}{dT}$ over $NB$ is dominant in Eq.~\ref{Trans-Thomson}. In the case of a time-varying magnetic field, the overall value of $Th_N$ is notably positive and decreases to zero at higher temperatures. 
 Overall, these results reveal that the leading transport coefficients, such as $\sigma_{el}$ and $\sigma_{H}$, have relatively less sensitivity towards a time-varying magnetic field as compared to those of higher-order thermoelectric transport coefficients. This enhanced sensitivity arises from their dependence on temperature derivatives and mixed charge–heat transport terms, making them more responsive to the temporal evolution of the magnetic field.

\section{Summary}
\label{summary}
In heavy-ion collisions, temperature gradients between the central and peripheral regions of the created fireball, especially in the presence of a non-zero baryon chemical potential, can introduce thermoelectric effects within the hot hadronic medium. The presence of a magnetic field induces anisotropy in the system by breaking rotational symmetry, enhancing direction-dependent thermoelectric responses. In this work, for the first time, we estimate the magneto-Thomson and transverse Thomson coefficients using different HRG models in the presence of a magnetic field. These higher-order thermoelectric coefficients arise from the temperature dependence of the leading-order thermoelectric coefficients. The Thomson coefficient quantifies the net generation or absorption of heat in the medium when current flows in the medium in the presence of a temperature gradient. We have studied this phenomenon in the presence of an external magnetic field. Here, the baryon conservation is considered for the heat conduction in the hot hadronic medium. Using the Gibbs-Duhem relation, we relate the gradients in baryon chemical potential to those in temperature, revealing the role of enthalpy in modifying the thermoelectric coefficients. The net leading-order thermoelectric coefficient vanishes for mesons due to their charge symmetry; mesons still have a significant effect on the leading-order thermoelectric coefficients because of their contributions to enthalpy. Furthermore, we explore the effect of different values of external magnetic fields on these transport properties. It is observed that in the presence of a higher magnetic field ($eB = 1.0~m_\pi^2$), both the $Th_B$ and $Th_N$ show higher values as compared to lower values of magnetic field ($eB = 0.1~m_\pi^2$). It is also observed that for both the coefficients, the effect of different models is mostly significant at higher temperatures. The results for all four models almost overlap at the lower temperatures.  In the absence of a magnetic field, the magneto-Thomson coefficient $Th_B$ reduces to the Thomson coefficient $Th$. On the other hand, the transverse Thomson coefficient $Th_N$ completely vanishes in the absence of a magnetic field because of the vanishing Lorentz force. Recently, both phenomena have also been studied in Ref.~\cite{Singh:2025geq} for the QGP medium in detail using the quasi-particle approach. The study is extended to include a time-varying magnetic field with an exponential decay profile, allowing an observation of its impact on the dynamics of the hadronic medium. The electrical conductivity and related heat current integrals show that a decaying magnetic field leads to smoother and weaker transport responses compared to the constant magnetic field case. The magneto-Thomson and transverse Thomson coefficients further show that the higher-order thermoelectric transport is strongly influenced by magnetic field dynamics. Concisely, the higher-order thermoelectric coefficients exhibit a significantly stronger sensitivity to time-varying magnetic fields than the leading-order transport coefficients such as electrical conductivities and heat currents, highlighting the importance of magnetic field evolution in realistic heavy-ion collision dynamics.

In the domain of spintronics, which includes both the charge and spin degrees of freedom of electrons for next-generation information processing technologies, the Thomson coefficient plays a fundamental role in uncovering spin-dependent thermoelectric phenomena. A specialized subfield, spin-caloritronics, focuses on the influence of temperature gradients on spin currents and magnetization dynamics~\cite{STARKOV2020165949}. For instance, Ref.\cite{Liu:2020dxg} discusses the generation of spin currents via the spin Hall effect, wherein a transverse spin current is induced in response to an electric field, analogous to the conventional Hall effect. Notably, thermoelectric-induced electric fields \cite{Singh:2024emy} can also contribute to the spin Hall current, establishing a direct link between thermal transport and spin dynamics. Recent experimental studies have further revealed the existence of a transverse Thomson effect in semimetallic alloys~\cite{takahagi}, highlighting the complex coupling between thermal gradients, charge transport, and magnetization. This motivates exploration of similar phenomena in the hot and dense QCD medium, where effects such as spontaneous magnetization may give rise to an anomalous transverse Thomson effect, an area yet to be investigated in the context of heavy-ion collisions. Therefore, a comprehensive understanding of higher-order thermoelectric phenomena in the hot QCD medium is not only crucial for studying transport properties but also essential for studying emergent spin-related effects, including the spin Hall effect and spin polarization phenomena~\cite{Singha:2022rlu}.
Overall, this study presents a detailed investigation of the higher-order thermoelectric properties of hot and dense hadronic matter within different HRG frameworks, emphasizing the impact of hadronic interactions and magnetic fields.

\section*{Acknowledgement}
KS and KKP acknowledge the financial aid from the University Grant Commission, Government of India. The authors gratefully acknowledge the DAE-DST, Govt. of India funding under the mega-science project – “Indian participation in the ALICE experiment at CERN” bearing Project No. SR/MF/PS-02/2021-
IITI (E-37123). 


\vspace{10.005em}


\end{document}